\documentclass[11pt]{article}
\usepackage{epsfig}
\usepackage{color}
\usepackage{amsmath}
\usepackage{amssymb}
\makeatletter
\def\singlespace{\def\baselinestretch{1}\@normalsize}

\@addtoreset{equation}{section}
\renewcommand{\baselinestretch}{1.412}

\renewcommand{\theequation}{\arabic{section}.\arabic{equation}}

\newcommand{\diag}{\mbox{diag}}

\newcommand{\convD}{\stackrel{D}{\longrightarrow}}

\newcommand{\var}{\mbox{var}}

\newcommand{\bS}{\mbox{\bf S}}

\newcommand{\ARMSE}{\mbox{ARMSE}}
\newcommand{\ARMISE}{\mbox{ARMISE}}
\newcommand{\MISE}{\mbox{MISE}}
\newcommand{\MSE}{\mbox{MSE}}

\newcommand{\CV}{\mbox{CV}}

\newcommand{\RSS}{\mbox{RSS}}

\newcommand{\bX}{\mbox{\bf X}}
\newcommand{\bbeta}{\mbox{\boldmath{$\beta$}}}

\newcommand{\eeta}{\mbox{\boldmath{$\eta$}}}
\newcommand{\bOmega}{\mbox{\boldmath{$\Omega$}}}
\newcommand{\bzeta}{\mbox{\boldmath{$\zeta$}}}
\newcommand{\beps}{\mbox{\boldmath{$\epsilon$}}}

\newcommand{\etal}{{\em et al.}\/}

\newcommand{\bx}{\mbox{\bf x}}
\newcommand{\bc}{\mbox{\bf c}}
\newcommand{\bI}{\mbox{\bf I}}
\newcommand{\ba}{\mbox{\bf a}}
\newcommand{\be}{\mbox{\bf e}}
\newcommand{\bA}{\mbox{\bf A}}
\newcommand{\bW}{\mbox{\bf W}}
\newcommand{\bG}{\mbox{\bf G}}

\newcommand{\bg}{\mbox{\bf g}}
\newcommand{\bb}{\mbox{\bf b}}
\newcommand{\bN}{\mbox{\bf N}}
\newcommand{\bR}{\mbox{\bf R}}

\newcommand{\bE}{\mbox{\bf E}}
\newcommand{\bM}{\mbox{\bf M}}
\newcommand{\bC}{\mbox{\bf C}}
\newcommand{\bB}{\mbox{\bf B}}
\newcommand{\bv}{\mbox{\bf v}}

\newcommand{\T}{\mbox{\tiny T}}
\newcommand{\bK}{\mbox{\bf K}}

\newtheorem{Theorem}{Theorem}[section]

\newtheorem{Corollary}[Theorem]{Corollary}

\def\beginn{\begin{eqnarray*}}
\def\endn{\end{eqnarray*}}
\def\beginy{\begin{eqnarray}}
\def\endy{\end{eqnarray}}
\def\n{\nonumber}

\begin{document}
\title{\LARGE \bf
Factor Models for Asset Returns Based on Transformed Factors
      }
\author{
Efang Kong
\\
School of Mathematics, Statistics and Actuarial Science
\\
University of Kent, UK
\and
Jialiang Li
\\
Department of Statistics \& Applied Probability
\\
National University of Singapore, Singapore
\and
Wenyang Zhang
\footnote{The corresponding author, Department of Mathematics, University of
York, York, YO10 5DD, United Kingdom, Email: \texttt{wenyang.zhang@york.ac.uk}.}\\
Department of Mathematics
\\
The University of York, UK
}  
\maketitle

\begin{abstract}

The Fama-French three factor models are commonly used in the description of 
asset returns in finance.  Statistically speaking, the Fama-French three factor 
models imply that the return of an asset can be accounted for directly by the 
Fama-French three factors, i.e. market, size and value factor, through a linear 
function.  A natural question is:  would some kind of transformed Fama-French 
three factors work better than the three factors?  If so, what kind of 
transformation should be imposed on each factor in order to make the 
transformed three factors better account for asset returns?  In this paper, we 
are going to address these questions through nonparametric modelling.  We 
propose a data driven approach to construct the transformation for each factor 
concerned.  A generalised maximum likelihood ratio based hypothesis test is 
also proposed to test whether transformations on the Fama-French three factors 
are needed for a given data set.  Asymptotic properties are established to 
justify the proposed methods.  Intensive simulation studies are conducted to 
show how the proposed methods work when sample size is finite.  Finally, we 
apply the proposed methods to a real data set, which leads to some interesting 
findings.

\vspace{0.2in}   

{\small \bf KEY WORDS}:  Backfitting, factor models, generalised maximum 
likelihood ratio test, kernel smoothing, transformed factor.

{\small \bf SHORT TITLE}:  FM for Asset Returns.

\end{abstract}

\section{Introduction}
\label{intro}

\subsection{Preamble}
\label{prea}

During the past two decades, much literature is devoted to explore the common 
factors in asset returns, see Ang \etal (2006), Brennan \etal (1998), Davis 
\etal (2000), Fama (1998), Fama and French (1993, 1996, 2010, 2015), Petkova 
(2006), Vassalou and Xing (2004), and the references therein.  Among the 
existing factor models, the Fama-French three factor models (FFTFM) are the 
arguably most commonly used models, they play a very important role in asset 
pricing and portfolio management.  The application of the FFTFM in fact goes 
beyond finance.   Fan \etal (2008) apply the FFTFM to introduce a structure for 
high dimensional covariance matrices, which significantly improves the 
estimation of high dimensional covariance matrices.  Measuring conditional 
dependence is an important topic in statistics with broad applications 
including graphical models.  Based on the FFTFM, Fan \etal (2015) have proposed 
a new conditional dependence measure.  Making use of the idea of the FFTFM, 
Guo \etal (2016) have proposed a dynamic structure for high dimensional 
covariance matrices and constructed an estimation procedure for the high 
dimensional covariance matrices with such structure.

\subsection{Motivating questions}
\label{mot}

Statistically speaking, the FFTFM imply that the return of an asset can be 
accounted for directly by the Fama-French three factors, i.e. market (Rm-Rf), 
size (SMB) and value factor (HML), through a linear function, see Fama and
French (1993).  A natural question is:  would some 
kind of transformed Fama-French three factors work better than the three 
factors?  If so, what kind of transformation should be imposed on each factor 
in order to make the transformed three factors better account for asset 
returns?  We can go even further to ask:  whether the linearity assumed in the 
FFTFM always holds?

To give a strong motivation for the models we are going to propose and 
investigate in this paper, we first study a data set freely downloaded from 
Kenneth French's website 

{\tt
http://mba.tuck.dartmouth.edu/pages/faculty/ken.french/data\_library.html
}  

\begin{figure}[htbp]
\centerline{\psfig{figure=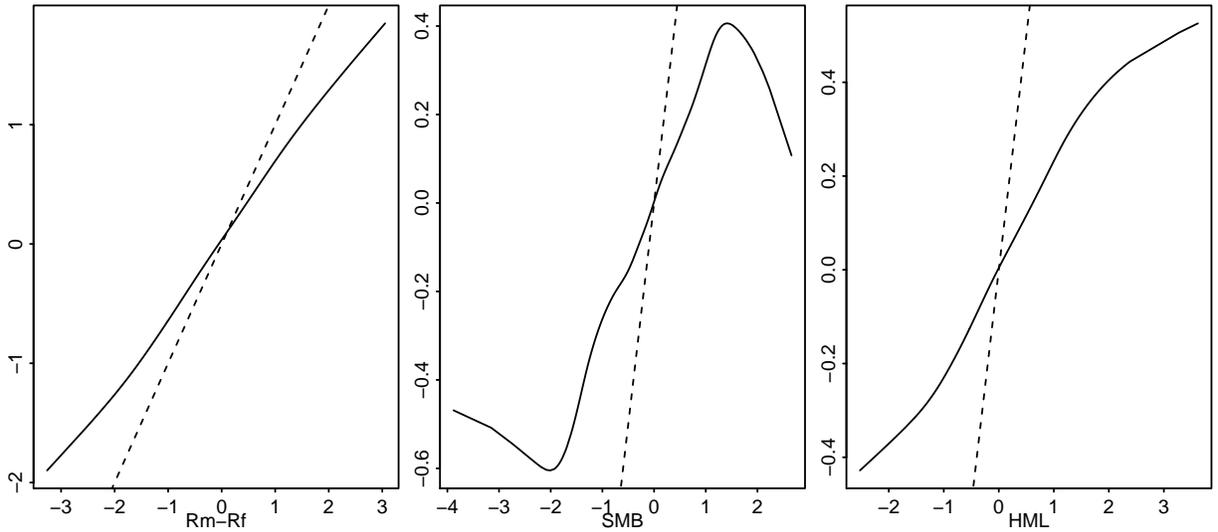,width=0.45\linewidth,angle=270}}
\caption[Figure 1]{\it The solid lines are the plots of the smoothed 
$\hat{\zeta}_{tk}$ against $x_{tk}$ for $k=1, \ 2, \ 3$, respectively.  This 
dashed lines are identity functions.}
\label{ff}
\end{figure}

The data set consists of the daily simple returns of $p_n = 49$ industry 
portfolios from 1927 to 2014.  Let $r_{tj}$ be the daily return of the $j$th 
portfolio at time $t$, $j=1, \ \cdots, \ 49$, $t=1, \ \cdots, \ T$, $x_{t1}$ 
(Rm-Rf), $x_{t2}$ (SMB), $x_{t3}$ (HML) be, respectively, the observations of 
the Fama-French three factors at time $t$.  For each given $j$, 
$j = 1, \ \cdots, \ 49$, we apply the FFTFM
\begin{equation}
r_{tj}
=
\alpha_j + \sum\limits_{k=1}^3 \beta_{jk} x_{tk} + \epsilon_{tj},
\quad
t = 1, \ \cdots, \ T; 
\label{FF}
\end{equation}
to fit $(r_{tj}, \ x_{t1}, \ x_{t2}, \ x_{t3})$, $t=1, \ \cdots, \ T$, and 
denote the obtained estimates of $\alpha_j$ and $\beta_{jk}$ as 
$\hat{\alpha}_j$ and $\hat{\beta}_{jk}$.  For each given $t$, 
$t=1, \ \cdots, \ T$, we conduct the following linear regression of 
$r_{tj} - \hat{\alpha}_j$ on 
$(\hat{\beta}_{j1}, \ \hat{\beta}_{j2}, \ \hat{\beta}_{j3})$
\begin{equation}
r_{tj} - \hat{\alpha}_j
=
\sum\limits_{k=1}^3 \hat{\beta}_{jk} \zeta_{tk} 
+ 
\varepsilon_{tj},
\quad
j = 1, \ \cdots, \ 49,
\label{FF1}
\end{equation}
and denote the estimates of $\zeta_{tk}$ as $\hat{\zeta}_{tk}$.  If the FFTFM 
were adequate, $\hat{\zeta}_{tk}$ would be a reasonably good estimate of 
$x_{tk}$, therefore, the plot of smoothed $\hat{\zeta}_{tk}$ against $x_{tk}$, 
$t=1, \ \cdots, \ T$, would be very close to an identity function for each 
given $k$.  Now, let's see whether this is the case.  For each $k$, 
we plot the smoothed $\hat{\zeta}_{tk}$ against $x_{tk}$ in Figure \ref{ff}.
It is clear the plot of smoothed $\hat{\zeta}_{tk}$ against $x_{tk}$ is not 
close to an identity function, which implies the FFTFM are not adequate, a 
transformation on each factor is necessary, and the transformed factors, 
denoted as $g_1(x_{t1})$, $g_2(x_{t2})$ and $g_3(x_{t3})$, would better account 
for asset returns than $x_{t1}$, $x_{t2}$ and $x_{t3}$ do.  Indeed, we will see 
this effect in the real data analysis section later on.  Now, the question 
is how to find the transformations $g_1(\cdot)$, $g_2(\cdot)$ and $g_3(\cdot)$.

\subsection{The proposed models}
\label{mod}

In order to find the transformations needed for the Fama-French three factors, 
we are going to propose a factor model based on transformed factors.  

Suppose we have $p$ factors, $x_1$, $\cdots$, $x_p$, and let 
$x_{t1}, \ \cdots, \ x_{tp}$ be 
the observations of the factors at time $t$, $t=1, \ \cdots, \ T$.  
Let $r_{tj}$ be the return of the $j$th asset at time $t$, $j=1, \ \cdots, 
\ n$, $t=1, \ \cdots, \ T$.  We assume 
\begin{equation}
r_{tj} 
= 
\alpha_j + \sum\limits_{k=1}^p \beta_{jk} g_k(x_{tk}) + \epsilon_{tj},
\quad
t = 1, \ \cdots, \ T; \ \ 
j = 1, \ \cdots, \ n,
\label{model}
\end{equation}
where $\alpha_j$, $\beta_{jk}$, and $g_k(\cdot)$, $j=1, \ \cdots, \ n$;  
$k=1, \ \cdots, \ p$, are unknown to be estimated, and
$$
E(\epsilon_{tj}|x_{t1}, \cdots, x_{tp}) = 0,
\quad
\var(\epsilon_{tj}|x_{t1}, \cdots, x_{tp}) = \sigma^2.
$$
It is clear (\ref{model}) is not identifiable.  To make (\ref{model}) 
identifiable, we assume 
\begin{equation}
g_k(x_{1k}) = x_{1k}
\mbox{ and }
E\{g_k(x_k)\} = 0,
\quad
k = 1, \ \cdots, \ p.
\label{iden}
\end{equation}

Model (\ref{model}) together with the identification condition (\ref{iden}) is 
the model we are going to address in this paper.  To connect the proposed model 
to the motivating questions, the $g_k(\cdot)$ in (\ref{model}) is the 
transformation needed for the $k$th factor.

There is fundamental difference between the proposed model (\ref{model}) and 
the additive models for panel data, which is the model (\ref{model}) with 
$\beta_{jk} g_k(x_{tk})$ being replaced by a completely unknown function 
$G_{jk}(x_{tk})$.  From statistical modelling point of view, the proposed model is more parsimonious, this is because there are only $p$ unknown functions and 
$(p+1)n$ unknown parameters in the proposed model, whilst there are $(p+1)n$ 
unknown functions in the additive models for panel data.  Most importantly, 
the proposed model (\ref{model}) is more meaningful, this is because from 
finance point of view, $g_k(x_{tk})$, $k=1, \ \cdots, \ p$, in (\ref{model}) 
act as common risk factors, whilst $G_{jk}(x_{tk})$, $j=1, \ \cdots, \ n$, 
$k=1, \ \cdots, \ p$, in the additive models depend on individual asset, 
therefore cannot be viewed as common risk factors.

The rest of the paper is organized as follows.  We begin in Section \ref{est} 
with a description of the estimation procedure for the unknowns in 
(\ref{model}).  Hypothesis test about whether a transformation is needed for 
each factor is discussed in Section \ref{test}.  Section \ref{asy} is devoted 
to the asymptotic properties of the proposed estimators and the hypothesis 
test.  Simulation studies are conducted in Section \ref{sim} to show how 
accurate the proposed estimators are and how powerful the proposed hypothesis 
test is when sample size is finite.  Finally, in Section \ref{real}, we apply 
the proposed modelling, estimation procedure and hypothesis test to the real 
data set mentioned in Section \ref{mot}, and some interesting findings will be 
presented.

\section{Estimation procedure}
\label{est}

In this section, we are going to construct the estimation procedure for the 
unknowns in (\ref{model}).  We are going to address the estimation of 
$g_k(\cdot)$s first, then $\alpha_j$s and $\beta_{jk}$s.

With a little bit abuse of notation, from now on, for any random error appears 
in a synthetic model in this section, we use $e_{tj}$ to denote, 
therefore, it may be different at different places.

\subsection{Estimation of $g_k(\cdot)$}
\label{est-g}

Let $G_{jk}(x_{tk}) = \beta_{jk} g_k(x_{tk})$, and re-write (\ref{model}) as
$$
r_{tj}
=
\alpha_j + \sum\limits_{k=1}^p G_{jk}(x_{tk}) + \epsilon_{tj},
\quad
t = 1, \ \cdots, \ T; \ \
j = 1, \ \cdots, \ n.
$$
For each given $j$, $j = 1, \ \cdots, \ n$, we apply the backfitting algorithm 
to estimate $G_{jk}(x_{tk})$, which is detailed as follows:  Let 
\beginy
\hat{\alpha}_j = \frac{1}{T} \sum\limits_{t=1}^T r_{tj}\label{ea}
\endy
and iterate the following two steps until convergence
\begin{enumerate}
\item Given the current $\tilde{G}_{jk}(x_{tk})$, $k=1, \ \cdots, \ p$.  For 
each $l$, $l=1, \ \cdots, \ p$, we run the following synthetic univariate 
nonparametric regression
$$
r_{tj} - \hat{\alpha}_j - \sum\limits_{k=1}^{l-1} \hat{G}_{jk}(x_{tk}) - 
\sum\limits_{k=l+1}^p \tilde{G}_{jk}(x_{tk})
=
G_{jl}(x_{tl}) + e_{tj},
\quad
t = 1, \ \cdots, \ T
$$
by the local linear modelling, which is detailed as follows.  For any given 
$u$, by the Taylor's expansion, we have 
$$
G_{jl}(x_{tl})
\approx
G_{jl}(u) + \dot{G}_{jl}(u) (x_{tl} - u)
$$
when $x_{tl}$ is in a small neighbourhood of $u$.  This leads to the 
following objective function for the local least squares estimation
\begin{equation}
\sum\limits_{t=1}^T
\left\{
r_{tj} - \hat{\alpha}_j - \sum\limits_{k=1}^{l-1} \hat{G}_{jk}(x_{tk}) -
\sum\limits_{k=l+1}^p \tilde{G}_{jk}(x_{tk})
-
c_{jl} - d_{jl} (x_{tl} - u)
\right\}^2
K_h(x_{tl} - u),
\label{llse}
\end{equation}
where $K_h(\cdot) = K(\cdot/h)/h$, $h$ is a bandwidth, $K(\cdot)$ is a kernel 
function, usually taken to be Epanechnikov kernel.  Minimise (\ref{llse}) with 
respect to $(c_{jl}, \ d_{jl})$, and denote the 
minimiser as $(\hat{c}_{jl}, \ \hat{d}_{jl})$.  The local linear estimator of 
$G_{jl}(u)$ is taken to be $\hat{c}_{jl}$, and denoted by 
$\check{G}_{jl}(u)$.  By simple calculation, we have
$$
\check{G}_{jl}(u)
=
(1, \ 0) \left(
\bOmega_l(u)^{\T} \bW_{l,h}(u) \bOmega_l(u)
\right)^{-1}
\bOmega_l(u)^{\T} \bW_{l,h}(u) \eeta_{jl},
$$
where $\bW_{l,h}(u)
=
\diag\left(K_h(x_{1l} - u), \ \cdots, \ K_h(x_{Tl}  - u) \right),
$
$$
\bOmega_l(u)
=
\left(
\begin{array}{cc}
1 & x_{1l} - u
\\
\vdots & \vdots
\\
1 & x_{Tl} - u
\end{array}
\right),
\quad
\eeta_{jl}
=
\left(
\begin{array}{cc}
r_{1j} - \hat{\alpha}_j - \sum\limits_{k=1}^{l-1} \hat{G}_{jk}(x_{1k}) -
\sum\limits_{k=l+1}^p \tilde{G}_{jk}(x_{1k})
\\
\vdots
\\
r_{Tj} - \hat{\alpha}_j - \sum\limits_{k=1}^{l-1} \hat{G}_{jk}(x_{Tk}) -
\sum\limits_{k=l+1}^p \tilde{G}_{jk}(x_{Tk})
\end{array}
\right).
$$
For each $x_{tl}$, the centralised $\check{G}_{jl}(x_{tl})$, denoted by 
$\hat{G}_{jl}(x_{tl})$, is
$$
\hat{G}_{jl}(x_{tl})
= 
\check{G}_{jl}(x_{tl}) - \frac{1}{T}\sum\limits_{t=1}^T \check{G}_{jl}(x_{tl}).
$$

\item Let $\tilde{G}_{jk}(x_{tk})$ be $\hat{G}_{jl}(x_{tk})$, and go to step 1.
\end{enumerate}
The iteration can be started by setting
$$
\tilde{G}_{jk}(x_{tk}) = 0, \ \ k=1, \ \cdots, \ p.
$$
With the final backfitting estimators $\hat{G}_{jl}(.)$s, we can construct the 
estimators of the functions $g_k(.)$s evaluated at the observation points as 
\beginy
\bar{g}_k(x_{tk}) 
= x_{1k} 
\frac{1}{n} \sum\limits_{j=1}^n 
\hat{G}_{jk}(x_{tk}) / \hat{G}_{jk}(x_{1k}),
\quad 
k=1, \, \cdots, \, p, \ t=1, \ \cdots, \ T.\label{rq}
\endy
For each $k$, $k=1, \ \cdots, \ p$, and any given $u$, viewing 
$\bar{g}_k(x_{tk})$ as a response variable, $x_{tk}$ as a covariate, we have 
the following synthetic univariate nonparametric regression model
\begin{equation}
\bar{g}_k(x_{tk})
=
g_k(x_{tk}) + e_{tk},
\quad
t = 1, \ \cdots, \ T.
\label{syn}
\end{equation}
Applying the local linear modelling to (\ref{syn}), similar to what we have 
done in step 1 in the backfitting algorithm for estimating $G_{jk}(x_{tk})$, 
we get an estimator of $g_k(u)$ 
$$
\hat{g}_k(u) 
=
(1, \ 0) \left(
\bOmega_k(u)^{\T} \bW_{k, \tilde h}(u) \bOmega_k(u)
\right)^{-1}
\bOmega_k(u)^{\T} \bW_{k, \tilde h}(u) \bzeta_{k},
\quad
\bzeta_{k}
=
\left(
\bar{g}_k(x_{1k}), \ \cdots, \ \bar{g}_k(x_{Tk})
\right),
$$
where $\tilde h$ is a bandwidth.  $\hat{g}_k(u)$ is our estimator of $g_k(u)$.

\subsection{Estimation of $\beta_{jk}$}
\label{est-a}

Estimates $\hat{\alpha}_j$, $j=1, \ \cdots, \ n$, from (\ref{ea}) and 
$\bar{g}_k(x_{tk})$, $t=1, \ \cdots, \ T$, $k=1, \ \cdots, \ p$, from 
(\ref{rq}) 
are plugged into (\ref{model}) as substitutes for their corresponding true but 
unknown counterparts so that we have the following synthetic linear model
\begin{equation}
r_{tj}
=
\hat\alpha_j + \sum\limits_{k=1}^p \beta_{jk} \bar{g}_k(x_{tk}) + e_{tj},
\quad
t = 1, \ \cdots, \ T.
\label{lin}
\end{equation}
Let $\bbeta_j = (\beta_{j1}, \ \cdots, \ \beta_{jp})^\top$.  We use the least 
squares estimator $\hat{\bbeta}_j$ of $\bbeta_j$ to estimate $\bbeta_j$, 
which is 
\begin{equation}
\hat{\bbeta}_j
=\left(
\bar{\bg}^{\T} \bar{\bg}
\right)^{-1}
\bar{\bg}^{\T} \bR_j,
\label{texas}
\end{equation}
where
$$
\bar{\bg}
=
\left(
\begin{array}{cccc}
  \bar{g}_1(x_{11}) & \cdots & \bar{g}_p(x_{1p})
\\
  \vdots & \ddots & \vdots
\\
 \bar{g}_1(x_{T1}) & \cdots & \bar{g}_p(x_{Tp})
\end{array}
\right) 
\ \mbox{and }\ 
\bR_j = (r_{1j}, \ \cdots, \ r_{Tj})^{\T}.
$$

\section{Hypothesis test}
\label{test}

In this section, we are going to address whether or not a transformation on 
each factor is significantly needed for a given data set.  We fomulate this 
question to a hypothesis test problem with null hypothesis 
\begin{equation}
H_0: \ g_1(x) = \cdots = g_p(x) = x.
\label{hyp}
\end{equation}
and alternative hypothesis being that transformations on the factors are 
needed.

Our hypothesis test is based on the generalised maximum likelihood ratio test, 
see Fan \etal (2001).  To construct the hypothesis test statistic, we first 
compute the residual sum of squares of the model (\ref{model}) under null 
hypothesis (\ref{hyp}).  Under the null hypothesis (\ref{hyp}), (\ref{model}) 
becomes the following linear model
\begin{equation}
r_{tj}
=
\alpha_j + \sum\limits_{k=1}^p \beta_{jk} x_{tk} + \epsilon_{tj},
\quad
t = 1, \ \cdots, \ T; \ \
j = 1, \ \cdots, \ n.
\label{null}
\end{equation}
Let 
$$
\bX
=
\left(
\begin{array}{cccc}
1 & x_{11} & \cdots & x_{1p}
\\
\vdots & \vdots & \ddots & \vdots
\\
1 & x_{T1} & \cdots & x_{Tp}
\end{array}
\right)
$$
By some simple calculations, we have the residual sum of squares of 
(\ref{null}) 
$$
\RSS_0
=
\sum\limits_{j=1}^n 
\bR_j^{\T}
\left\{
\bI_T - \bX (\bX^{\T} \bX)^{-1} \bX^{\T}
\right\}
\bR_j,
$$
where $\bI_T$ is an identity matrix of size $T$.

On the other hand, the residual sum of squares of (\ref{model}) is
$$
\RSS_1
=
\sum\limits_{j=1}^n \sum\limits_{t=1}^T
\left(
r_{tj}
-
\hat{\alpha}_j 
- 
\sum\limits_{k=1}^p \hat{\beta}_{jk} \bar{g}_k(x_{tk})
\right)^2.
$$
Based on the idea in Fan \etal (2001), we propose the following test statistic 
for the null hypothesis (\ref{hyp})
$$
\lambda 
= 
\frac{nT}{2} \frac{\RSS_0 - \RSS_1}{\RSS_1}.
$$
We reject $H_0$ when $\lambda > c$, where $c$ is determined by 
$$
P(\lambda > c | H_0) = \alpha, 
$$
$\alpha$ is the significant level.

In the implementation of the proposed hypothesis test, the distribution of 
$\lambda$ under null hypothesis can be either estimated by bootstrap or 
approximated by its asymptotic distribution presented in Section \ref{asy}.

\section{Asymptotic properties}
\label{asy}

For each $k$, $k=1, \ \cdots, \ p$, as far as the estimator of $g_k(u)$ is 
concerned, because the theoretical properties of $\hat{g}_k(u)$ easily follow 
from those of $\bar{g}_k(x_{tk})$ at the expense of further cumbersome 
notations, we only present the asymptotic properties of $\bar{g}_k(x_{tk})$.

For simplicity, we assume that observation points all lie in the interior of 
the support of $\bx $ and focus on local polynomial fittings of odd degrees, as 
the expressions become considerably more complicated with boundary points or 
in the case of even degrees (Opsomer  and Ruppert, 1997).
Write $\beps_j=(\epsilon_{1j},\cdots,\epsilon_{\T j})^\top,\,j=1,\cdots,n.$
Then regarding the estimates discussed in Section \ref{est}, we have
\begin{Theorem}   
\label{theorem1}
Under the Assumptions given in the Appendix, 
\begin{description}
\item{(1)} $\bar g_k(x_{tk})=
g_k(x_{tk})+\gamma_{tk}^\top \bS_{k}\frac{1}{n}
\sum\limits_{j=1}^n\beta^{-1}_{jk}\beps_j +o_p(T^{-1/2})
$   uniformly with respect to $t=1, \ \cdots, \ T$ and $k=1, \ \cdots, \ p$. 
\item{(2)}$T^{1/2}(\hat{\alpha}_j - \alpha_j) \convD N(0, \ \sigma^2)$
 
\item{(3)}$\hat{\bbeta}_{j}-\bbeta_j
=c_0(K)\frac{1}{Tn}\sum\limits_{j'=1}^n \bA_{j'|j}\beps_j+o_p(T^{-1/2})$.
\end{description}
\end{Theorem}
Definitions of $T\times 1$ vector $\gamma_{tk}$, $T\times T$ matrix $\bS_{k}$, constant $c_0(K)$ 
and $p\times T$ matrix $\bA_{j'|j}$ are given in the Appendix.
It easily follows that $\hat g_k(.)$ converges at a nonparametric rate of $(Th_k)^{-1/2}$.

Let $R(K)=\int  K^2(u)du$. For the testing statistic in Section \ref{test}, we have; 
\begin{Theorem} 
\label{theorem2} Suppose conditions in Theorem \ref{theorem1} hold, and for ease of exposition, $h_1=h_2=\cdots=h_p=h$.  Then 
under the null hypothesis (\ref{hyp}), 
$$
P\{\sigma_{\T}^{-1}[\lambda-npK(0)h^{-1}]<t\}
\longrightarrow
\Phi(t),\quad \mbox{when }  {T\to\infty},
$$
where $\Phi(\cdot)$ is the standard normal distribution function, 
$$
\sigma^2_T
=
\sigma^4R(K)h^{-1}\Big\{ \sum\limits_{j,k}  
c_k\{4+\sum\limits_{j'\ne j} (\beta_{jk}/\beta_{j'k})^2\}
+n(n-1) \sum\limits_{k=1}^p c_k\Big\}.
$$
\end{Theorem}
Constant $c_k$ is to be defined in the Appendix. 

Theorem \ref{theorem2} provides us the asymptotic distribution of the proposed 
test statistic for the null hypothesis (\ref{hyp}), which can be used to 
estimate the critical value of the proposed hypothesis test in 
Section \ref{test}.

\section{Simulation studies}
\label{sim}

In this section, we are going to use a simulated example to demonstrate how 
accurate the proposed estimators are.  We will also examine the power of the 
proposed hypothesis test for the null hypothesis (\ref{hyp}).  As the 
asymptotic distribution of the test statistic involves unknown parameters and 
some constants which are hard to calculate, we will use bootstrap approach to 
compute the critical value for the test.

We generate data according to model (\ref{model}).  Specifically, each element 
of $X_t=(x_{t1}, \ \cdots, \ x_{tp})^T$ is independently generated from a 
uniform distribution over $[-1, \ 1]$, and each random error $\epsilon_{tj}$ is 
generated from $N(0, \ 1)$.  We set $p=4$ and 
\begin{equation}
\begin{array}{c}
g_1(x_1)=\sin(2.5\pi x_1), \quad g_2(x_2)=x_2^3, 
\quad g_3(x_3)=\sin(0.5\pi x_3),
\\[3mm]
g_4(x_4)
= 
\left[1/\left\{1+\exp(-x_4)\right\}-0.5\right]/\left\{1/(1+e^{-1})-0.5\right\}.
\end{array}
\label{setg}
\end{equation}
We will consider various $n$ and $T$ in our simulation study.  For each $n$ 
and $T$, the interecepts $\alpha_j$s in the model (\ref{model}) are 
independently generated from $N(3, \ 0.5)$ and the slopes $\beta_{jk}$s are 
independently generated from $N(3.5, \ 0.5)$.  Once these $\alpha_j$s and 
$\beta_{jk}$s are generated, we fix them across all simulations for the given 
$n$ and $T$.

Let $\MSE(\hat{\alpha}_j)$ and $\MSE(\hat{\beta}_{jk})$ be the mean squared 
errors of $\hat{\alpha}_j$ and $\hat{\beta}_{jk}$, respectively.  We use 
$\ARMSE_{\alpha}$ and $\ARMSE_{\beta}$, which are defined as
$$
\ARMSE_{\alpha}
=
\frac{1}{n} \sum\limits_{j=1}^n 
\left\{\alpha_j^{-2}\MSE(\hat{\alpha}_j)\right\},
\quad
\ARMSE_{\beta}
=
\frac{1}{np} \sum\limits_{j=1}^n \sum\limits_{k=1}^p 
\left\{\beta_{jk}^{-2} \MSE(\hat{\beta}_{jk})\right\},
$$
to assess the accuracy of our estimation for the interecepts $\alpha_j$s and 
for the slopes $\beta_{jk}$s, respectively.  Let $\MISE_k$ be the mean 
integrated squared error of $\hat{g}_k(\cdot)$.  We use $\ARMISE$, which is 
defined as
$$
\ARMISE
=
\frac{1}{p} \sum\limits_{k=1}^p 
\MISE_k \left\{\int g_k( u )^2 du \right\}^{-2} 
$$
to assess the accuracy of our estimation for the unknown functions 
$g_k(\cdot)$s.

We consider various $n$ and $T$.  For each given $n$ and $T$, we do 500 
simulations, the obtained $\ARMSE_{\alpha}$ and $\ARMSE_{\beta}$ are presented 
in Table \ref{simtab1}, and the obtained $\ARMISE$ is reported in 
\ref{simtabl2}.  The two tables show our estimation procedure works very well.

\begin{table}[htbp]
\begin{center}
\caption{\bf 
The Performance of Our Estimation for Unknown Parameters
} 
\label{simtab1}
\vspace{0.2in}
\begin{tabular}{lc|ccc}
\hline\hline
& & $T=200$ & $T=800$ & $T=1500$\\
\hline
$n=20$ & $\ARMSE_{\alpha}$ & .0136 & .0031 & .0017\\
& $\ARMSE_{\beta}$ & .1328& .0083 & .0023\\
\hline
$n=50$ & $\ARMSE_{\alpha}$ & .0143 & .0029 &  .0019\\
& $\ARMSE_{\beta}$ & .4915 & .0110 & .0005\\
\hline
$n=80$ & $\ARMSE_{\alpha}$ & .0105& .0028& .0018\\
& $\ARMSE_{\beta}$ & .0166& .0102 & .0016\\
\hline
\hline
\end{tabular}
\end{center}
\end{table}

\begin{table}[htbp]
\begin{center}
\caption{\bf The $\ARMISE$s of Our Estimation for Unknown Functions} 
\label{simtabl2}
\vspace{0.2in}
\begin{tabular}{l|ccc}
\hline\hline
&  $T=200$ & $T=800$ & $T=1500$\\
\hline
$n=20$ & .2302 & .0749 & .0008 \\
\hline
$n=50$ & .1879 & .0486 & .0003 \\
\hline
$n=80$ & .0361 & .0165 & .0004 \\
\hline\hline
\end{tabular}
\end{center}
\end{table}

We now examine how powerful the proposed hypothesis test is.  To evaluate the 
performance of the proposed hypothesis test, we use the same data generating 
setting as described earlier and only modified the true functional forms of the 
factors to be
$$
{\bf g}
=
\rho \left(g_1(x_1), \ g_2(x_2), \ g_3(x_3), \ g_4(x_4) \right)^{\T}
+(1-\rho) {\bf x},
\quad
\bx = (x_1, \ x_2, \ x_3, \ x_4)^{\T}
$$ 
where each $g_k(\cdot)$ was given as in (\ref{setg}).  When $\rho=0$, the null 
hypothesis (\ref{hyp}) is true.  When $\rho$ is away from zero, the true 
functional forms of the factors are not identity functions, and we should 
reject the null hypothesis (\ref{hyp}).

\begin{figure}[h]
\begin{center}
\includegraphics[scale = 0.4, angle = 0]{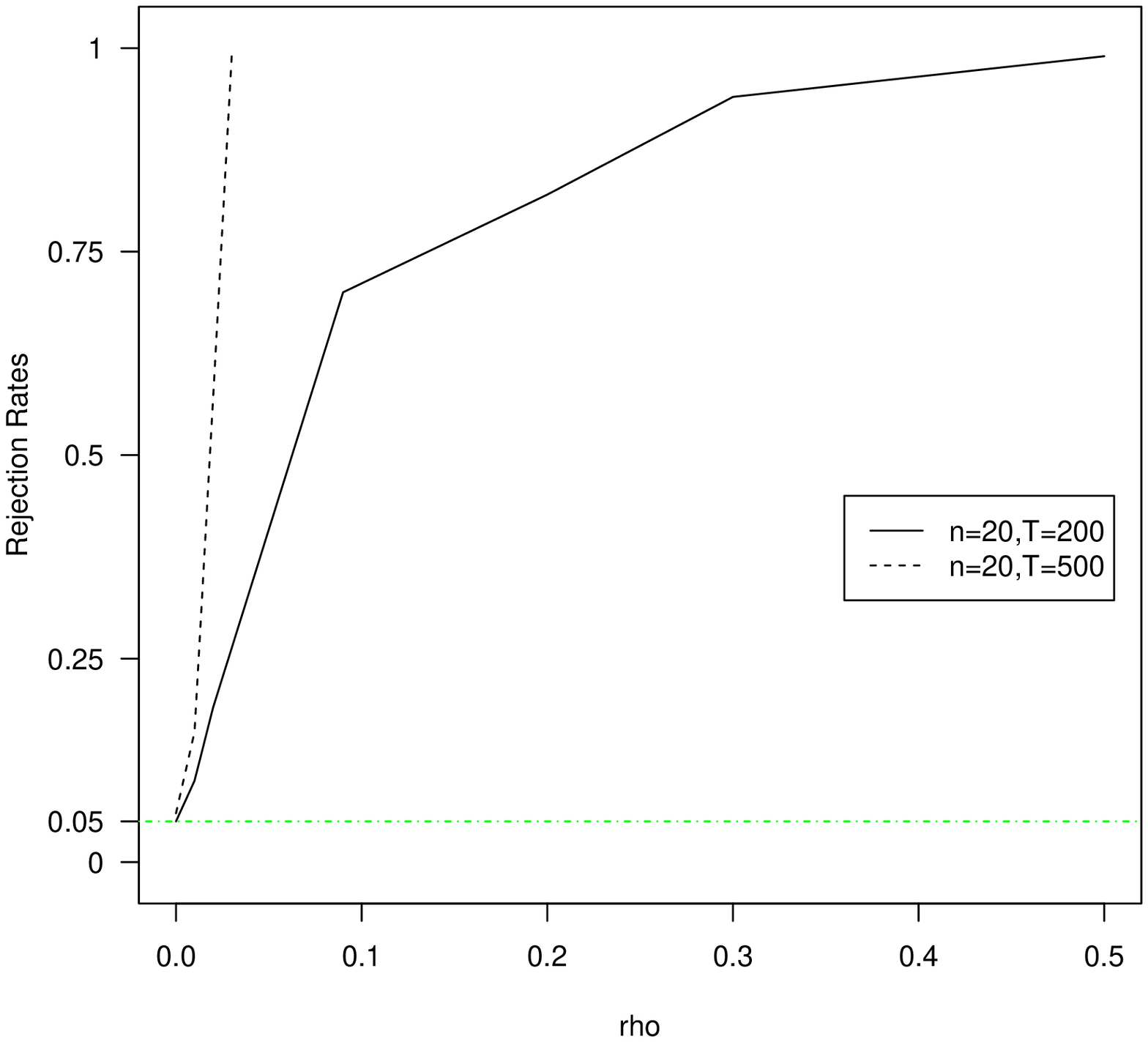}
\includegraphics[scale = 0.4, angle = 0]{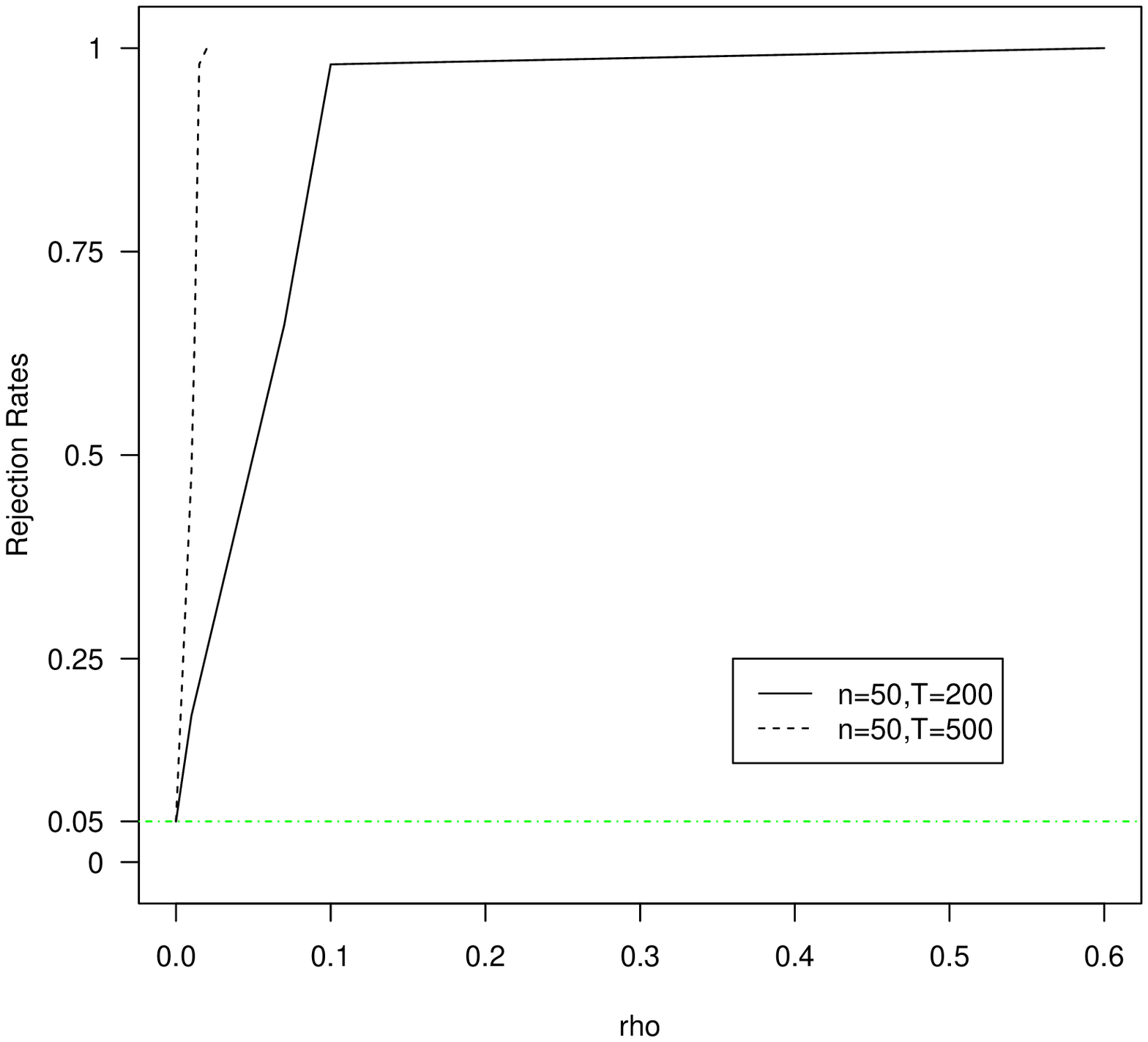}
\vspace{-6pt}
\caption{The power function of the proposed test when $n=20$ or $50$, and 
$T=200$ or $T=500$.}
\label{simfig}
\end{center}
\end{figure}

We set the significance level to be $0.05$, and consider the power function of 
the proposed test for various $n$ and $T$.  For each given $\rho$, $n$ and $T$, 
we do $500$ simulations.  In each simulation, we generate a data set and apply 
the proposed hypothesis test to the generated data to test the null hypothesis 
(\ref{hyp}).  The critical value is computed through a bootstrap sample, of 
size $1000$, of the test statistic $\lambda$ under null hypothesis.  The value 
of the power function at $\rho$ is defined as the rejection rate of the test 
among the $500$ simulations, and actual size of the test is the value of the 
power function at $\rho = 0$.  The obtained power function is reported in 
Figure \ref{simfig} for various $n$ and $T$, and the actual size is reported 
in Table \ref{size}.  Taking the Monte Carlo error, which is of size 
$(0.05\times 0.95/500)^{1/2} \approx 0.01$, into account, we can safely 
conclude that the actual size of our test is very close to $0.05$ based on 
Table \ref{size}.  Figure \ref{simfig} shows the rejection rates approach one 
as $\rho$ becomes large, indicating that our test has high power to reject the 
null when it is false.  In general, the test performance improves as $n$ and 
$T$ increase.

\begin{table}[htbp]
\begin{center}
\caption{\bf The Actual Size of the Proposed Test}
\label{size}
\vspace{0.2in}
\begin{tabular}{l|cc}
\hline\hline
&  $T=200$ & $T=500$\\
\hline
$n=20$ & 0.053 & 0.052 \\
\hline
$n=50$ & 0.055 & 0.048 \\
\hline\hline
\end{tabular}
\end{center}
\end{table}

\section{Real data analysis}
\label{real}

In this section, we apply the proposed methods to the data set mentioned in 
Section \ref{mot}.  We will show the transformations on the Fama-French three 
factors are significantly necessary for this data set by the proposed 
hypothesis test, and construct the transformation needed for each factor
by the proposed estimation method.  We will also show how much improvement the
proposed transformation can result in, in terms of accounting for the return
of an asset.

\begin{figure}[htbp]
\centerline{\psfig{figure=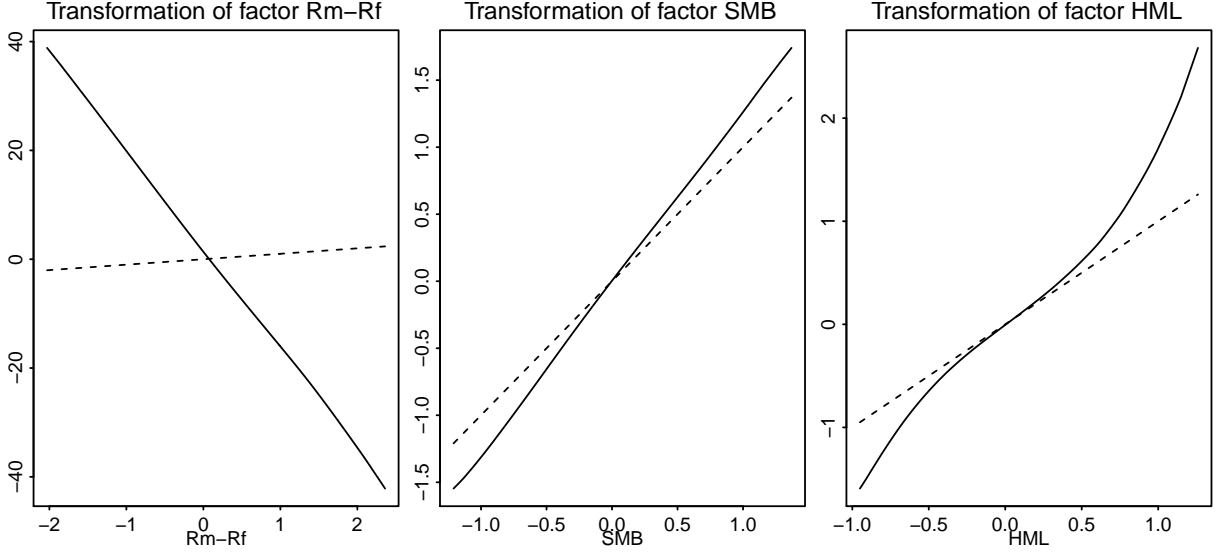,width=0.45\linewidth,angle=270}}
\caption[Figure 1]{\it The solid lines are functions $\hat{g}_k(\cdot)$, 
$k=1, \ 2, \ 3$, respectively.  This dashed lines are identity functions.}
\label{real1}
\end{figure}

To investigate whether the FFTFM (\ref{FF}) is appropriate for this data set, we consider fitting the proposed model (\ref{model}) to the data set. 
The estimated functions $\hat{g}_k(\cdot)$ for the three factors were plotted 
in Figure \ref{real1}. 

Figure \ref{real1} shows clearly $\hat{g}_k(\cdot)$, $k=1, \ 2, \ 3$, are not 
identity functions, and $\hat{g}_3(\cdot)$ is even not a linear function.  
Indeed, when applying the proposed hypothesis test to this data set to test the 
null hypothesis (\ref{hyp}), we obtain a p-value of $0.003$, suggesting that 
the null hypothesis should be rejected.  The p-value is computed through a 
bootstrap sample, of size $1000$, of the test statistic $\lambda$ under null 
hypothesis.  We therefore conclude that it is necessary to make a 
transformation on each of the three factors in the FFTFM.

\begin{figure}[htbp]
\centerline{\psfig{figure=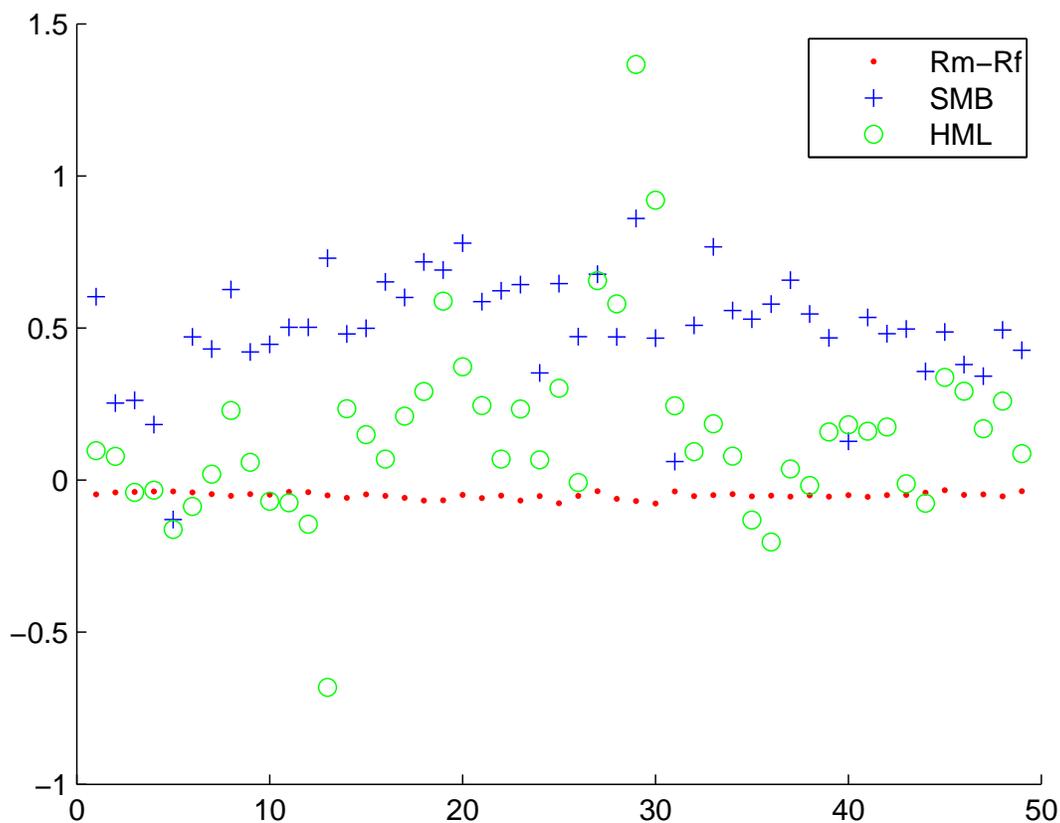,width=6.5in}}
\caption{\it Estimated coefficients. }
\label{port2}
\end{figure}

The estimated coefficients of the three transformed factors, $g_k(x_k)$, 
$k=1, \ 2, \ 3$, for all $n=49$ portforlios are shown in Figure \ref{port2}.  
The coefficients for the transformed Rm-Rf, $g_1(x_1)$, were mostly negative 
and very close to -0.05. The coefficients for the transformed SMB, $g_2(x_2)$, 
are mostly positive around 0.50 and much greater than those for the transformed 
Rm-Rf. The coefficients for the transformed HML, $g_3(x_3)$, are not so 
homogeneous and may be quite different for the individual portforlios.

We now investigate how much improvement the transformed common factors can 
make in terms of accounting for the return of an asset.

For a given model, let $E_{ji}$ be the squared prediction error of the 
prediction for the simple return of the $j$th portfolio on the $i$th day from 
the last, based on this model and the observations before the $i$th day from 
the last.  
We construct the cross-validation sum for this model based on the prediction 
errors for the last $30$ days, and define it as
$$
\CV 
= 
\frac{1}{30 \times 49} \sum\limits_{i=1}^{30} \sum\limits_{j=1}^{49} E_{ji}.
$$
We compute, respectively, the CVs for the FFTFM and the proposed model 
({\ref{model}), and find the ratio of CV of the FFTFM to the CV of the proposed 
model is 1.3587.   This indicates the proposed model can make more than $35\%$ 
improvement in terms of accounting for the return of an asset.

\section*{References}
 
\begin{description}

\item Buja, A., Hastie, T. and Tibshirani, R. (1989) Linear smoothers and additive models. {\em Ann. Statist} 17,    543--555.
\item Fan, J. and Jiang, J. (2005) Nonparametric inferences for additive models. {\em JASA}  {\bf 100}, 890-902.


\item Le Cam, L. and Yang, G. (1990)  {\em Asymptotic in Statistics: Some Basic Concepts.} New York: Springer-Verlag. 
\item Masry, E. (1996) Multivariate regression estimation
Local polynomial fitting for time series.
{\em Stochastic Processes and their Applications} 65, 81--101.

\item Masry, E. (1996) Multivariate local polynomial regression for time series: uniform strong consistency and rates. {\em Journal of Time Series Analysis} 17, 571–-599.

\item Opsomer, J. (2000) Asymptotic Properties of Backfitting Estimators. {\em Journal of Multivariate Analysis} 73, 166–-179.

\item Opsomer, J. and Ruppert, D. (1997) Fitting a bivariate additive model by local polynomial regression. {\em   Ann. Statist}  25,   186--211.
\item Yu, B. (1994) Rates of convergence for empirical processes of stationary mixing sequences. {\em Annals of probability} 22, 94-116.
\item Ang, A., Hodrick, R. J., Xing, Y. and Zhang, X. (2006).  The 
cross‐section of volatility and expected returns.  {\it The Journal of 
Finance}, {\bf 61}, 259-299.

\item Brennan, M. J., Chordia, T. and Subrahmanyam, A. (1998).  Alternative 
factor specifications, security characteristics, and the cross-section of 
expected stock returns.  {\it Journal of Financial Economics}, {\bf 49}, 
345-373.

\item Davis, J. L., Fama, E. F. and French, K. R. (2000).  Characteristics, 
covariances, and average returns: 1929 to 1997.  {\it The Journal of Finance}, 
{\bf 55}, 389-406. 

\item Fama, E. F. (1998).  Market efficiency, long-term returns, and behavioral 
finance.  {\it Journal of Financial Economics}, {\bf 49}, 283-306.

\item Fama, E. F. and French, K. R. (1993).  Common risk factors in the returns 
on stocks and bonds.  {\it Journal of Financial Economics}, {\bf 33}, 3-56. 

\item Fama, E. F. and French, K. R. (1996).  Multifactor explanations of asset 
pricing anomalies.  {\it The Journal of Finance}, {\bf 51}, 55-84.

\item Fama, E. F. and French, K. R. (2010).  Luck versus skill in the 
cross-section of mutual fund returns.  {\it The Journal of Finance}, {\bf 65}, 
1915-1947.

\item Fama, E. F. and French, K. R. (2015). A five-factor asset pricing model.  
{\it Journal of Financial Economics}, {\bf 116}, 1-22.

\item Fan, J., Fan, Y. and Lv, J. (2008). High dimensional covariance matrix
estimation using a factor model.  {\it J. Econometrics}, {\bf 147}, 186-197.

\item Fan, J., Feng, Y. and Xia, L. (2015). A conditional dependence measure 
with applications to undirected graphical models.  arXiv:1501.01617.

\item Fan, J., Zhang, C. and Zhang, J. (2001).  Generalized likelihood 
ratio statistics and Wilks phenomenon.  {\it Ann. Statist.} {\bf 29}, 
153-93.

\item Guo, S., Box, J. and Zhang, W. (2016).  A dynamic structure for high 
dimensional covariance matrices and its application in portfolio allocation.  
{\it Journal of the American Statistical Association}, to appear.

\item Petkova, R. (2006).  Do the Fama–French factors proxy for innovations in 
predictive variables?  {\it The Journal of Finance}, {\bf 61}, 581-612.

\item Vassalou, M. and Xing, Y. (2004).  Default risk in equity returns.  {\it 
The Journal of Finance}, {\bf 59}, 831-868.

\end{description}

\section*{Appendix}
\setcounter{equation}{0}
\renewcommand{\theequation}{A.\arabic{equation}}
\setcounter{equation}{0}
\renewcommand{\theequation}{A.\arabic{equation}}

It is clear from the estimation  procedure as described in Section \ref{est-g} that the statistical properties of the estimated   component functions $g_k(.)$ as well as  those of $\hat\alpha_j,\, \hat\beta_{jk}$  could only be derived based on the asymptotics concerning the backfitting estimators $\hat{G}_{jk}(.)$.    To present the relevant results on this aspect,
we need to  introduce more notations. 
Let   $f(.)$ be the joint density function of $(x_{t1},\cdots,x_{tp})$, and $f_k(.),\, k=1,\cdots,p$, the marginal
density of the $k$th covariate $x_{tk}$. Denote by $f_{l;k}(.,.)$, the joint density of $x_{tk}$ and 
$x_{(t+l)k}$;   $f_{l;k,k'}(u,u,v,v)$, the joint pdf 
of $x_{tk},x_{(t+l)k},x_{tk'},x_{(t+l)k'}$ evaluated at  $(u,u,v,v).$
For any $l\ge 1$, $k,k'=1,\cdots,p, k\ne k',$ define
\beginn
a_{l;k}=\int \frac{f_{l;k}(u,u)}{f^2_{k}(u)}du,
\  
b_{l;k,k'}=\int \frac {f_{l;k,k'}(u,u,v,v)}{f_k(u)f_{k'}(v)}dudv.
\endn
We assume that
\beginn
c_k:=\lim\limits_{T\to \infty}\Big|\frac{1}{T^2}\sum\limits_{l=1}^{T-1}(T-l)a_{k,l}\Big|<\infty, \quad
\lim\limits_{T\to \infty}\sup\limits_{k\ne k'}\Big|\frac{1}{T^2}\sum\limits_{l=1}^{T-1}(T-l)b_{l;k,k'}\Big|<\infty;
\endn

The following conditions are assumed throughout of the paper.
\begin{description}
\item {[A1]} $\bx_t=(x_{t1},\cdots,x_{tp})^\top$ is a $p-$variate  stationary processes
and is strongly mixing, i.e. 
\beginn
\gamma[\iota]:= \sup\limits_{\overset{A\in {\mathbf{F}}_{-\infty }^{0}}{B\in {\mathbf{F}}%
_{\iota}^{\infty }}}|P[AB]-P[A]P[B]|\rightarrow 0,\ \mbox{as }%
k\rightarrow \infty ,
\endn
where  ${\mathbf{F}}_{s_1}^{s_2}$ is the $\sigma -$ algebra of events generated
by  $\{\bx_t:s_1\leq t\leq s_2\} $ and $\gamma[\iota]$ is referred to as the strong mixing coefficient.
Moreover, $\sum\limits_{\iota=1}^\infty \iota^a \gamma [ \iota]^{1-2/v}<\infty$  for some $v>2$ and $a>1-2/v$.

\item {[A2]} The kernel function $K(.)$ is bounded and continuous with a compact support; its first order
derivative has a finite number of sign changes over its support.

\item {[A3]} Both the joint $f(.)$ and the marginal densities  $f_k(.),\, k=1,\cdots,p$ are   bounded and continuous with compact support; their first order
derivatives also have a finite number of sign changes over their supports.

\item{[A4]} $\sup\limits_{u,u'}|f_{l;k}(u,u')-f_k(u)f_k(u')|\le A_1<\infty$ for all $l\ge 1$.

\item {[A5]} As $T\to \infty,$ $h_k\to 0, \ Th_k/\log n\to \infty,\ Th_k^{\iota_k+2}\to 0$ for all $ k=1,\cdots,p$.

\item {[A6]} There exists a sequence $v_n$ of positive integers satisfying $v_{\T}\to\infty$ and $v_{\T}=o((nh)^{1/2})$ such that
$(T/h)^{1/2}\gamma[v_{\T}]\to 0$ as $T\to\infty. $

\end{description}
Assumption [A1] is relevant since the backfitting estimator $\hat G_{jk}(.)$ in this paper is built on dependent observations, $\{r_{tj}, \, t=1,\cdots,T\}$, which is different from the set-up in  Opsomer (2000) with independent observations. Strongly mixing could be replaced by a weaker condition, such as $\beta-$mixing or even $\phi-$mixing, but in that case additional requirement on these alternative mixing coefficients will then be necessary; see e.g. Masry (1996).
[A2] could be relaxed to allows kernel functions of unbounded support provided that $u^{\iota_k+1}K(u)\to 0$ as $u\to\infty.$

For   $l=0,1,\cdots,$
write the  $lth $ moment of the kernel function $K(.)$ as $\mu_l(K)
:=\int u^lK(u)du$  and   $R_l=\int u^l K^2(u)du,\ $ and $R(K)=R_0$. 
For $k=1,\cdots,p,$ let $g^{(\iota)}_{k}(.)$ denote the $\iota$th derivative
 of component function $g_k(.)$, and  write
\beginn
  {\bg}^{(\iota)}_{k}=
\left[\begin{array}{c}
  g^{(\iota)}_{k}(x_{1k})\\
\vdots\\
  g^{(\iota)}_{k}(x_{\T k})
\end{array}\right],
\quad
E({\bg}^{(\iota)}_{k}|\bX^k)=
\left[\begin{array}{c}
E(  g^{(\iota)}_{k}(x_{ik})|x_{1k}) \\
\vdots\\
E(  g^{(\iota)}_{k}(x_{ik})|x_{\T k})
\end{array}\right],\ k\ne k'.
\endn

The backfitting algorithm described in Section \ref{est-g} is based on local linear smoothing. Here we give a more general results on backfitting estimators based on local polynomial smoothing where functions $g_k(.)$ are  locally approximated by a  polynomial of degree $\iota_k$, $k=1,\cdots,p$. Define the following smoother matrix for the $k$th component function:
\beginy
\bS_{k}=(\bS_{k,x_{1k}},\cdots,\bS_{k,x_{\T k}})^\top,\label{smooth}
\endy
where $\bS_{k,u}$ represents the transpose of the equivalent kernel for the $k$th covariate at the point $u$:
\beginn
\bS_{k,u}=\bK_{k }(u)\bX_{k}(u)\Big[\bX_{k }(u)^\top  \bK_{k }(u)\bX_{k }(u)\Big]^{-1}\be_{1k}^\top,
\endn
 $\be_{1k}$ is the $(\iota_k+1)\times 1$ vector with a one in the first position and zeros elsewhere,
\beginn
\bX_{k}(u)=\left[\begin{array}{cccc}
1& x_{1k}-u&\cdots& (x_{1k}-u)^{\iota_k} \\
\vdots&\vdots&\ddots&\vdots\\
1& x_{\T k}-u&\cdots& (x_{\T k}-u)^{\iota_k}
\end{array}\right],\quad  \bK_{k }(u)=\diag\left(K_h(x_{1k} - u), \ \cdots, \ K_h(x_{Tk}  - u) \right).
\endn
Further define the centered smoothing matrix 
$\bS_k^*=(\bI-{\bf 1}_{\T}{\bf 1}_{\T}^\top)\bS_k $,
   $\bW_{[-k]}$, the smoother matrix for the $(p-1)$-variate function $G_{j}^{(-k)}(.)=\sum_{l=1,l\ne k}^p G_{jl}(.)$, and  \
   $\bG_{jk}=({G}_{jk}(x_{1k}),\cdots,{G}_{jk}(x_{\T k}))^\top$,
      the vector of the $k$th 
 component function evaluated at the observation points. 
Then regarding $\hat \bG_{jk}$, the backfitting estimator of $\bG_{jk}$, we have

\begin{Corollary}\label{c1}  Given $\bX,$  the
   conditional bias and variance of $ \hat{\bG}_{jk},\,  j=1,\cdots, n, \ k=1,\cdots,p,$ are respectively
\beginn
&&\hspace{-0.7cm}E(\hat{\bG}_{jk}- {\bG}_{jk}|\bX)=(\bI-\bS_k^*\bW_{[-k]})^{-1}
\Big[\frac{1}{(\iota_k+1)!}h^{\iota_k+1}\mu_{\iota_k+1}(K)\beta_{jk}\Big({\bg}^{(\iota_k+1)}_{k}
-E({\bg}^{(\iota_k+1)}_{k})\Big)-\bS_k^*\bB_{j[-k]}\Big]\\
&&\hspace{2.5cm}+O_p(T^{-1/2})+o_p(h^{\iota_k+1}),\\
&&Var(\hat{\bG}_{jk}(x_{tk})|\bX)=   \{nhf_k(x_{tk})\}^{-1}R_K\sigma^2+o_p((nh)^{-1},
)\endn
 where
 \beginn
 \bB_{j[-k]}=E\Big(\bW_{[-k]}(\bR_j-\bG_{jk})|\bX\Big)-\sum\limits_{l=1:l\ne k}^p \bG_{jl}.
 \endn
 \end{Corollary}
The bias expression in Corollary \ref{c1} is still a recursive formula, and as commented in Opsomer (2000),
a non-recursive asymptotic bias expression can be derived, but the expressions become very complicated even for $p=3.$
Nevertheless, the order of the asymptotic bias could be easily decided for any $p$:
\beginn
E(\hat{\bG}_{jk}- {\bG}_{jk}|\bX)=O_p(\sum\limits_{k=1}^p h_k^{\iota_k+1}).
\endn
Apparently, if  $g_k(.), \, k=1,\cdots,p$ are all smooth enough,  and with polynomial fitting of  high enough $\iota_k$
degrees employed,   this bias term  could be made relatively negligible compared to asymptotic stochastic error.
 We will make use of this fact in later sections in the asymptotic  study of  $ \hat g_k(.),$ and $\hat\bbeta_j$.

We now move on to prove Theorem \ref{theorem1}, starting with  more notations.
Let $$
c_0(K)
=
\sum\limits_{\iota=0}^{\iota_k}[\bN^{-1}]_{(\iota+1)1}\mu_\iota(K), 
$$ 
where $\bN$ represents the $(\iota_k+1)\times (\iota_k+1)$ matrix, whose $(i,j)$th element is  $\mu_{i+j-2}(K)$, and $[\bN^{-1}]_{(\iota+1)1}$ stands for the $(\iota+1,1)$th element of its inverse matrix.
Define the $T\times 1$ vectors
$$\gamma_{tk}=(-g_k(x_{tk}),0,\cdots,0,1,0,\cdots,0)^\top,\quad t=2,\cdots, T$$
with $1$ as the $t$th entry.
For any given $k,k'=1,\cdots,p$, define    
$$c_{k,k'}(u)=E[g_k(x_{tk})|x_{tk'}=u],\quad \bc_{k,k'}=[c_{k,k'}(x_{1k'}),\cdots,c_{k,k'}(x_{\T k'})]^\top,$$ 
$$
\bA_{j'|j}=[\ba_{1j'|j},\cdots,\ba_{pj'|j}]^\top,
\quad
\ba_{kj'|j}=\sum\limits_{k'=1}^p \frac{\beta_{jk'}}{\beta_{j'k'}}\bc_{k,k'}, 
\ \ j,j'=1, \ \cdots, \ n; \ k=1, \ \cdots, \ p.
$$

  {\bf Proof of Theorem \ref{theorem1}  } Similar computations as in the proof of the second assertion of Corollary \ref{c1} lead to
  \beginn
 \hat{\bG}_{jk}-E\hat{\bG}_{jk}= \bS_{k}\beps_j+O_p(T^{-1/2}),\quad j=1,\cdots,n,\ k=1,\cdots,p,
 \endn
uniformly in  over all elements of the matrices; see, also Opsomer (2000, pp. 178).
For ease of exposition,  write the asymptotic bias and stochastic error of  $\hat{\bG}_{jk}$ as
\beginn
\bb_{jk}=E\hat{\bG}_{jk}-{\bG}_{jk}\equiv (b_{jk,1},\cdots,b_{jk,\T})^\top,
\quad \bv_{jk}= \bS_{k}\beps_j\equiv (v_{jk,1},\cdots,v_{jk,\T})^\top .
\endn
As a result, we have
\beginn
&&\frac{\hat{G}_{jk}(x_{tk})}{\hat{G}_{jk}(x_{1k})}
=\frac{\beta_{jk} g_k(x_{tk})+b_{jk,t}+v_{jk,t}}{\beta_{jk} +b_{jk,1}+v_{jk,1}}\\
&&\quad= g_k(x_{tk})+\frac{b_{jk,t}}{\beta_{jk}}+\frac{v_{jk,t}}{\beta_{jk}}
-\frac{g_k(x_{tk})b_{jk,1}}{\beta_{jk}}-\frac{g_k(x_{tk})v_{jk,1}}{\beta_{jk}}
+o_p(h_k^{\iota_k+1}+T^{-1/2}). 
\endn
Since without loss of generality, 
 we could always assume that $x_{1k}=1$ and whence for each $t=2, \,\cdots, \, T,$
\beginy
\n\bar g_k(x_{tk})&=&\frac{1}{n}\sum\limits_{j=1}^n\hat{G}_{jk}(x_{tk})/\hat{G}_{jk}(x_{1k}) \\
\n &=& g_k(x_{tk})+\frac{1}{n}\sum\limits_{j=1}^n\Big(\frac{b_{jk,t}}{\beta_{jk}}-\frac{g_k(x_{tk})b_{jk,1}}{\beta_{jk}}\Big)
\\
&&+\frac{1}{n}\sum\limits_{j=1}^n\Big(\frac{v_{jk,t}}{\beta_{jk}}
-\frac{g_k(x_{tk})v_{jk,1}}{\beta_{jk}}\Big)
+o_p(h_k^{\iota_k+1}+T^{-1/2}),\label{shen}
\endy
again uniformly in $t$ and $k$.

Since the second (bias) term on the RHS of (\ref{shen}) is of order $o(T^{-1/2})$ if  $g_k(.)$ is smooth enough and a  large enough $\iota_k$ is used,    we have
\beginn
 \bar g_k(x_{tk})=g_k(x_{tk})+\gamma_{tk}^\top \bS_{k}\frac{1}{n}\sum\limits_{j=1}^n\beta^{-1}_{jk}\beps_j +o_p(T^{-1/2}).
\endn
Since $\beps_j, \,j=1,\cdots,n$ are all iid errors with zero mean and variance $\sigma^2$,
the asymptotic variance of $ \hat g_k(x_{tk})$ is such that
\beginy
\Big(n^{-2}\sum\limits_{j=1}^n\sigma^2_j/\beta^2_{jk}\Big)\gamma_{tk}^\top \bS_{k}\bS_{k}^\top\gamma_{tk} .\label{way}
\endy
Using standard results in polynomial smoothing (Masry, 1996) that
\beginy
&&[\bS_{k}]_{ij}=\{f_k(x_{ik})\}^{-1}\frac{1}{Th_k}
\sum\limits_{\iota=0}^{\iota_k}[\bN^{-1}]_{(\iota+1)1}\Big(\frac{x_{jk}-x_{ik}}{h_k}\Big)^{\iota} K\Big(\frac{x_{jk}-x_{ik}}{h_k}\Big)
\label{trioka}.
\endy
 Consequently
\beginn
&&\hspace{-0.5cm} [\bS_{k}\bS_{k}^\top]_{ii'}=\{f_k(x_{ik})f_k(x_{i'k})\}^{-1}\frac{1}{T^2h^2_k}\sum\limits_{j=1}^{T}\Big\{\sum\limits_{\iota=0}^{\iota_k}[\bN^{-1}]_{(\iota+1)1}\Big(\frac{x_{jk}-x_{ik}}{h_k}\Big)^{\iota} K\Big(\frac{x_{jk}-x_{ik}}{h_k}\Big)\Big\}\\&&\hspace{7cm}\times
\Big\{\sum\limits_{\iota=0}^{\iota_k}[\bN^{-1}]_{(\iota+1)1}\Big(\frac{x_{jk}-x_{i'k}}{h_k}\Big)^{\iota} K\Big(\frac{x_{jk}-x_{i'k}}{h_k}\Big)\Big\}\\
&&\hspace{1cm}=\{f_k(x_{ik})f_k(x_{i'k})\}^{-1}\frac{1}{Th_k}\sum\limits_{\iota,\iota'=0}^{\iota_k}[\bN^{-1}]_{(\iota+1)1}[\bN^{-1}]_{(\iota'+1)1}R(i,i';\iota,\iota')+O_p((Th_k)^{-3/2})
\endn
where
$$R(i,i';\iota,\iota')=\int \Big(\frac{x_{ik}-x_{i'k}}{h_k} +t\Big)^{\iota'}t^\iota K(t)K(s+t)dt.$$ Therefore,
\beginn
&& \gamma^\top_{tk} \bS_{k} =([\bS_{k}]_{tj}-g_k(x_{tk})*[\bS_{k}]_{1j})=O((Th_k)^{-1})\\
&&\gamma_{tk}^\top \bS_{k}\bS_{k}^\top\gamma_{tk} =\{g_k(x_{tk})\}^2[\bS_{k}\bS_{k}^\top]_{11}-2[\bS_{k}\bS_{k}^\top]_{1t}g_k(x_{tk})+[\bS_{k}\bS_{k}^\top]_{tt}
\endn
This together with (\ref{way}) implies that the asymptotic variance of   $ \hat g_k(x_{tk})$ is of order $O((Th_k)^{-1/2})$. 

As for the estimates of the parameters, first note that the results on 
 $\hat\alpha_j$ easily follow  from  (\ref{model}), (\ref{iden}) and the strong mixing conditions [A1].
To examine the asymptotic properties of  $\hat\beta_{jk}$, {\it least square estimate} (\ref{texas}) derived from model (\ref{lin}), first note that
according to Theorem \ref{theorem1}, we have  that
\beginy
\bar{\bg}= {\bg}+O_p((Th_k)^{-1/2}), \quad (\frac{1}{T}\bar{\bg} ^\top \bar{\bg})^{-1}=\Sigma_g^{-1}+O_p((Th_k)^{-1/2}),\label{rm}
\endy
 uniformly in all elements of the matrix, where 
$${\bg}=\left[\begin{array}{ccc}
g_1(x_{11})&\cdots&g_p(x_{1p})  \\
 {g}_1(x_{21})&\cdots&{g}_p(x_{2p})\\
\vdots&\vdots&\vdots\\
{g}_1(x_{\T 1})&\cdots&{g}_p(x_{\T p})
\end{array}\right]=\left[\begin{array}{ccc}
1&\cdots&1  \\
 {g}_1(x_{21})&\cdots&{g}_p(x_{2p})\\
\vdots&\vdots&\vdots\\
{g}_1(x_{\T 1})&\cdots&{g}_p(x_{\T p})
\end{array}\right],
$$
since without loss of generality, 
 we have  assumed that $x_{1k}=1$ whence $g_k(x_{1k})=x_{1k}=1$.
These, together with the decomposition
$R_j=\hat\alpha_j {\bf 1}_{\T}+(\alpha_j-\hat\alpha_j){\bf 1}_{\T}+\hat{\bg}{\bbeta}_{j}+({\bg}-\hat{\bg}){\bbeta}_{j} +\beps_j$ and the root-$T$ consistency of $\hat\alpha_j$,
lead to
\beginn
\hat{\bbeta}_{j}&=&(  \bar{\bg} ^\top \bar{\bg})^{-1}\bar{\bg}^\top (R_j-\hat\alpha_j {\bf 1}_{\T})\\
&=&{\bbeta}_{j}+(  \bar{\bg} ^\top \hat{\bg})^{-1}\bar{\bg}^\top(\alpha_j-\hat\alpha_j){\bf 1}_{\T}+( {\bar\bg} ^\top {\hat\bg})^{-1}\bar{\bg}^\top({\bg}-\bar{\bg}){\bbeta}_{j}
+( {\bar\bg} ^\top {\bar\bg})^{-1}\bar{\bg}^\top({\bg}-\hat{\bg}){\beps}_{j}\\
&=&{\bbeta}_{j}+( {\bg} ^\top {\bg})^{-1} {\bg}^\top(\alpha_j-\hat\alpha_j){\bf 1}_{\T}+( {\bg} ^\top {\bg})^{-1}{\bg}^\top({\bg}-\bar{\bg}){\bbeta}_{j}+o_p(T^{-1/2})\\
&=&{\bbeta}_{j}+\Sigma_g^{-1}T^{-1}{\bg}^\top({\bg}-\bar{\bg}){\bbeta}_{j}+\Sigma_g^{-1}T^{-1}\bg\beps_j+o_p(T^{-1/2})
\endn
where  we've used the following facts:
$$T^{-1}{\bg}^\top{\bf 1}_{\T}=O_p(T^{-1/2}),\quad T^{-1}({\bg}-\bar{\bg})  \beps_{j}=O_p(T^{-1/2}).$$
This means the error arisen from the pre-estimation of $\alpha_j$ has been `averaged out' and thus of no impact.
To show that $\hat\bbeta_j$ is asymptotically normal, first note that the  $k$th element of ${\bg}^\top({\bg}-\bar{\bg}){\bbeta}_{j}$
is given by
\beginn
&&\frac{1}{n}\sum\limits_{j'=1}^n\sum\limits_{k'=1}^p \frac{\beta_{jk'}}{\beta_{j'k'}}\Big[\sum\limits_{t=2}^T g_{k}(x_{tk})\gamma_{tk'}\Big]^\top\bS_{k'}\beps_{j'}
\quad k=1,\cdots, p;\quad \mbox{with}\\
&&\sum\limits_{t=2}^T g_{k}(x_{tk})\gamma_{tk'}=\Big[-\sum\limits_{t=2}^T g_{k}(x_{tk})g_{k'}(x_{tk'}),\, g_{k}(x_{2k}),\cdots,g_{k}(x_{Tk})\Big]^\top.
\endn
Therefore,
\beginn
& \Big[\sum\limits_{t=2}^T g_{k}(x_{tk})\gamma_{tk'}\Big]^\top\bS_{k'}=c_0(K)
 \bc_{k,k'} ^\top+O_p((Th_k)^{-1/2})&\\
& \frac{1}{n}\sum\limits_{j'=1}^n\sum\limits_{k'=1}^p \frac{\beta_{jk'}}{\beta_{j'k'}}\Big[\sum\limits_{t=2}^T g_{k}(x_{tk})\gamma_{tk'}\Big]^\top\bS_{k'}\beps_{j'}
=c_0(K)
\frac{1}{n}\sum\limits_{j'=1}^n\Big[\sum\limits_{k'=1}^p \frac{\beta_{jk'}}{\beta_{j'k'}}\bc_{k,k'}\Big]^\top\beps_{j'}+o_p(T^{1/2}).&
\endn
Since $\beps_{j'},\,j'=1,\cdots,n$ are independent $MN(0,\bI_T)$, the asymptotic normality of $T^{1/2}(\hat\bbeta_j-\bbeta_j)$ thus follows with asymptotic
 variance given by
\beginn
c^2_0(K)\Sigma_g^{-1}n^{-2}\Big(\sum\limits_{j'=1}^n T^{-1}\bA_{j'|j} \bA^\top_{j'|j}\Big)\Sigma_g^{-1},
\endn
which is finite.  $\hspace{\fill}\square$


\noindent{\bf Proof of Theorem \ref{theorem2}} 
First of all, it is easy to see that $\RSS_{1}/(nT)\to \sigma^2$ in probability as $T\to \infty.$
So we just need to concern us with the numerator $\RSS_{0}-\RSS_{1}=\sum\limits_{j=1}^n \RSS_{0,j}-\RSS_{1,j},$
where
 \beginn
&&\RSS_{1,j}
= 
R^\top_{j}[\bI_T-\tilde {\bg}(\tilde{\bg}^\top \tilde{\bg})^{-1}
\tilde{\bg}^\top]R^\top_{j};
\quad
\tilde{\bg}=
\left(
\begin{array}{cccc}
1&  \bar{g}_1(x_{11}) & \cdots & \bar{g}_p(x_{1p})
\\
 \vdots &  \vdots & \ddots & \vdots
\\
1& \bar{g}_1(x_{T1}) & \cdots & \bar{g}_p(x_{Tp})
\end{array}
\right)
\\
&&\RSS_{0,j}= R^\top_{j}[\bI_T-\bX(\bX^\top \bX)^{-1}\bX^\top]R^\top_{j}
= \beps^\top_{j}[\bI_T-\bar\bX(\bar\bX^\top \bar\bX)^{-1}\bar\bX^\top]\beps^\top_{j}.
\endn
Note that the second identity follows from the fact that   $\bX(\bX^\top\bX)^{-1}\bX^\top$ is invariant if 
 ${\bX}$ is replaced with ${\bX}$ right-multiplied   by a diagonal matrix and that
$$\bar\bX=\left[\begin{array}{cccc}
1&1&\cdots&1  \\
1& {g}_1(x_{21})&\cdots&{g}_p(x_{2p})\\
\vdots&\vdots&\vdots&\vdots\\
1&{g}_1(x_{\T 1})&\cdots&{g}_p(x_{\T p})
\end{array}\right]=\bX\left[\begin{array}{ccccc}
 1&x^{-1}_{11}&0&\cdots&0  \\
0& 0 & x^{-1}_{12}&0& 0\\
\vdots&\vdots&\vdots&\vdots&\vdots\\
0&0 &\cdots&0&x^{-1}_{1 p} 
\end{array}\right].
$$
With a slight abuse of notation, we revert to the old notation of $\bg$ in place of $\bar\bX.$ 
Write $\tilde{\bg}=\bg+\delta$, $\Delta=\bg^\top\delta+\delta^\top\bg$, $\Gamma=({\bg}^\top {\bg})^{-1}\bg^\top$ so that
\beginn
&&\tilde{\bg}^\top \tilde{\bg}={\bg}^\top {\bg}+ {\bg}^\top \delta+\delta^\top  {\bg}+\delta^\top \delta,\\
&&(\tilde{\bg}^\top \tilde{\bg})^{-1}
=({\bg}^\top {\bg})^{-1}-({\bg}^\top {\bg})^{-1}\Delta({\bg}^\top {\bg})^{-1}+O_p((Th_k)^{-1}),\\
&&\tilde {\bg}(\tilde{\bg}^\top \tilde{\bg})^{-1}\tilde{\bg}^\top={\bg}({\bg}^\top {\bg})^{-1}{\bg}^\top
+\delta\Gamma+\Gamma^\top \delta^\top+\delta({\bg}^\top {\bg})^{-1}\delta^\top-\Gamma^\top\Delta\Gamma\\
&&\quad-\delta ({\bg}^\top {\bg})^{-1}\Gamma ({\bg}^\top {\bg})^{-1}\bg^\top
- \bg({\bg}^\top {\bg})^{-1}\Gamma ({\bg}^\top {\bg})^{-1}\delta^\top
- \delta({\bg}^\top {\bg})^{-1}\Gamma ({\bg}^\top {\bg})^{-1}\delta^\top+O_p((Th_k)^{-1}).
\endn
Since $R_j=\bg\bbeta_{j}+\beps_{j}$, we have the following partition of the difference of the two Residual Sum of Squares:
\beginy
\n&&\RSS_{0,j}-\RSS_{1,j}= -2R^\top_{j}\delta\Gamma R_{j}
+R^\top_{j}\Gamma^\top\Delta\Gamma R_{j}-{ R^\top_{j}\delta({\bg}^\top {\bg})^{-1}\delta^\top R_{j}}
\\
&&\hspace{3cm} +2R^\top_{j}\bg({\bg}^\top {\bg})^{-1}\Delta ({\bg}^\top {\bg})^{-1}\delta^\top R_{j}+ { R^\top_{j}\delta({\bg}^\top {\bg})^{-1}\Delta ({\bg}^\top {\bg})^{-1}\delta^\top R_{j}}.\label{mao}
\endy
We start with the third term on the RHS of (\ref{mao}), and will show that
\beginy
R^\top_{j}\delta({\bg}^\top {\bg})^{-1}\delta^\top R_{j}=o_p(h^{-1}).\label{porter}
\endy
Some useful results are
\beginy
&&E[\beps_j^\top \delta({\bg}^\top {\bg})^{-1}\delta^\top \beps_j]
=\frac{1}{T}E[\beps_j^\top\delta\Sigma_g^{-1}\delta^\top \beps_j](1+O_p(1))\le 
\frac{C}{T}E\|\delta^\top \beps_j\|^2=o(h_k^{-1})\label{jia}\\
\n&&E\|\delta^\top \beps_j\|^2\le p\max\limits_{k}
E\Big(\sum\limits_{t=2}^T [\gamma^\top_{tk} \bS_{k}\sum\limits_{j=1}^n\beta^{-1}_{jk}\beps_j ]\epsilon_{tj}\Big)^2
\\
\n&&E\Big(\sum\limits_{t=2}^T [\gamma^\top_{tk} \bS_{k}\sum\limits_{j=1}^n\beta^{-1}_{jk}\beps_j ]\epsilon_{tj}\Big)^2
=\sum\limits_{t=2}^T E[\gamma^\top_{tk} \bS_{k}\sum\limits_{j=1}^n\beta^{-1}_{jk}\beps_j ]^2\epsilon_{tj}^2\\
\n&&\hspace{5cm}+\sum\limits_{t\ne t'} E\Big([\gamma^\top_{tk} \bS_{k}\sum\limits_{j=1}^n\beta^{-1}_{jk}\beps_j ]
[\gamma^\top_{t'k} \bS_{k}\sum\limits_{j=1}^n\beta^{-1}_{jk}\beps_j ]\epsilon_{tj}\epsilon_{t'j}\Big)\\
\n&&\hspace{5cm}= O(T^2(Th)^{-2})= O( h^{-2}),
\endy
where   the last equality follows from the fact that
$\gamma^\top_{tk} \bS_{k} =([\bS_{k}]_{tj}-g_k(x_{tk})*[\bS_{k}]_{1j})=O((Th)^{-1})$.

(\ref{porter}) thus follows from (\ref{jia}), if we could also show that $\bbeta_j^\top\bg^\top\delta({\bg}^\top {\bg})^{-1}\delta^\top \bg\bbeta_j=O_p(1)$, which could be proved in a manner  similar to (\ref{jia}).  Specifically, it is obviously of the same order as $T^{-1}$ times the trace of $\bg^\top\delta\Sigma_g^{-1}\delta^\top \bg$, which in turn 
of the same order as the largest norm of the $p$ columns of  $\delta^\top \bg$:  
its $(k,l)$th element for any $l,k=1,\cdots,p,$ is given by
\beginn
&&\sum\limits_{t=2}^T \frac{x_{tl}}{x_{1l}} \Big(
\gamma_{tk}^\top \bS_{k}\frac{1}{n}\sum\limits_{j=1}^n\beta^{-1}_{jk}\beps_j\Big)
=\sum\limits_{j=1}^n \beta^{-1}_{jk}\beps^\top_j\bS^\top_{k}\Big(\sum\limits_{t=2}^T \frac{x_{tl}}{x_{1l}}  \gamma_{tk}\Big)
=O_p(1),
\endn
where the last equality follows from the following facts:
\beginn
&&\sum\limits_{t=2}^T \frac{x_{tl}}{x_{1l}}\gamma_{tk}
=\Big[-\sum\limits_{t=2}^T \frac{x_{kl}x_{tl}}{x_{kl}x_{1l}},\frac{x_{2l}}{x_{1l}},\cdots,\frac{x_{Tl}}{x_{1l}}\Big]^\top,\\
&&
\sum\limits_{t'=1}^T[\bS_{k}]_{t'j}\Big(\sum\limits_{t=1}^T \frac{x_{tl}}{x_{1l}}  \gamma_{tk}\Big)=O(1)+O_p((Th)^{-1/2}).
\endn

Next, we will show that for the last term on the RHS of (\ref{mao}) the following holds:
\beginy
R^\top_{j}\delta({\bg}^\top {\bg})^{-1}\Delta ({\bg}^\top {\bg})^{-1}\delta^\top R_{j}=O_p((Th)^{-1}).\label{parker}
\endy
{\ref{parker}}
This is based on the following identities:
\begin{description}
\item{(A)} $
 \beps^\top_{j}\delta({\bg}^\top {\bg})^{-1}\Delta ({\bg}^\top {\bg})^{-1}\delta^\top \beps_{j}=O_p((Th)^{-1})$;
 \item{(B)}  $\bbeta_j^\top\bg^\top\delta({\bg}^\top {\bg})^{-1}\bg^\top\delta ({\bg}^\top {\bg})^{-1}\delta^\top \bg\bbeta_j =O_p(T^{-2})$.
\end{description}
That (A) holds is argued as follows. Firstly $\beps^\top_{j}\delta({\bg}^\top {\bg})^{-1}\Delta ({\bg}^\top {\bg})^{-1}\delta^\top \beps_{j}
=2\beps^\top_{j}\delta({\bg}^\top {\bg})^{-1}\bg^\top\delta ({\bg}^\top {\bg})^{-1}\delta^\top \beps_{j} $,
and the $k$th ($k=1,\cdots,p$) element of $\beps^\top_{j}\delta $ is such that
\beginn
\sum\limits_{j'=1}^n \beta_{j'k}^{-1}\Big( \epsilon_{tj}\gamma_{tk}^\top \bS_{k}\Big)\beps_{j'}
&=&\sum\limits_{j'=1}^n \beta_{j'k}^{-1}\Big( 
 [-\sum\limits_{t=2}^T\frac{x_{tk}}{x_{1k}}\epsilon_{tj},\epsilon_{2j},\cdots,\epsilon_{Tj} ] \bS_{k}\Big)\beps_{j'}\\
 &=&\sum\limits_{j'=1}^n \beta_{j'k}^{-1}\Big[ 
 \sum\limits_{t=2}^T\epsilon_{tj} [\bS_{k}]_{t,t'}-\sum\limits_{t=2}^T\frac{x_{tk}}{x_{1k}}\epsilon_{tj}[\bS_{k}]_{1,t'},t'=1,\cdots,T\Big]\beps_{j'}.
\endn
Since $\sum\limits_{t=2}^T\epsilon_{tj} [\bS_{k}]_{t,t'}=O_p((Th)^{-1/2})$ and
 $\sum\limits_{t=2}^T\frac{x_{tk}}{x_{1k}}\epsilon_{tj}=O_p(T^{-1/2})$, uniformly in $t'=1,\cdots,T,$ 
 whence $\beps^\top_{j}\delta =O_p((T/h)^{1/2})$.

 We now move on to the second term on the RHS of (\ref{mao}): $R^\top_{j}\Gamma^\top\Delta\Gamma R_{j}$,
 which again is bounded by two times 
 \beginn
 \beps^\top_{j}\bg ({\bg}^\top {\bg})^{-1} \bg^\top\delta({\bg}^\top {\bg})^{-1}\bg^\top\beps_{j}
 +\bbeta^\top_{j}\bg^\top \bg ({\bg}^\top {\bg})^{-1} \bg^\top\delta({\bg}^\top {\bg})^{-1}\bg^\top\bg\bbeta_{j}=O_p(1),
 \endn
where for the last equality we used the fact that $\bg^\top\beps_{j}=O_p(T^{1/2}).$
 
 Now the only term left to be dealt with is $R^\top_{j}\delta\Gamma R_{j}$, which equates to
 \beginy
&&\n R^\top_{j}\delta\bbeta_j+R^\top_{j}\delta({\bg}^\top {\bg})^{-1} \bg^\top \beps_{j}  
 =\beps^\top_{j}\delta\bbeta_j+\bbeta_j^\top\bg^\top\delta\bbeta_j+\bbeta_j^\top {\bg}^\top \delta({\bg}^\top {\bg})^{-1} \bg^\top \beps_{j}\\&& \hspace{5cm}+
\beps_{j}^\top \delta({\bg}^\top {\bg})^{-1} \bg^\top \beps_{j};\   \label{divide}
 \endy
where $\bbeta_j^\top {\bg}^\top \delta({\bg}^\top {\bg})^{-1} \bg^\top \beps_{j}=O_p(T^{-1/2})$ and  
$\bbeta_j^\top\bg^\top\delta\bbeta_j=O_p(1)$. The $k$th element of  $\beps^\top_{j}\delta$:
\beginn
\sum\limits_{j'=1}^n\beta^{-1}_{j'k}\Big(\sum\limits_{t=2}^T \epsilon_{tj}\gamma_{tk}^\top\Big) \bS_{k}\beps_{j'}
=\sum\limits_{j'\ne j} \beta^{-1}_{j'k}\Big(\sum\limits_{t,t'=2}^T \epsilon_{tj}\epsilon_{t'j'} [\bS_{k}]_{t,t'} \Big)+
 \beta^{-1}_{jk}\Big(\sum\limits_{t,t'=2}^T \epsilon_{tj}\epsilon_{t'j} [\bS_{k}]_{t,t'} \Big),
\endn
has its mean given by
\beginy
&& \beta^{-1}_{jk}\sigma^2 \sum\limits_{t=2}^T  [\bS_{k}]_{t,t}  =  K(0)\beta^{-1}_{jk}\sigma^2h^{-1}(1+o_p(1));\label{tong}
\endy
and its  second moment  as \\
\beginn
&&\sigma^4 \sum\limits_{j'\ne j} \beta^{-2}_{j'k}\sum\limits_{t,t'=2}^T[\bS_{k}]^2_{t,t'}+\beta^{-2}_{jk} \mu_4\sum\limits_{t=2}^T  [\bS_{k}]^2_{t,t}\\
&&\hspace{2cm}+\beta^{-2}_{jk}
\sigma^4 \sum\limits_{t<t'}\{[\bS_{k}]^2_{t,t'}+[\bS_{k}]^2_{t',t}+2[\bS_{k}]_{t,t'}[\bS_{k}]_{t',t}+2[\bS_{k}]_{t,t}[\bS_{k}]_{t',t'}\} 
\\
&&=\sigma^4 \sum\limits_{j'\ne j} \beta^{-2}_{j'k}\sum\limits_{t,t'=2}^T[\bS_{k}]^2_{t,t'}+\beta^{-2}_{jk} (\mu_4-\sigma^4)\sum\limits_{t=2}^T  [\bS_{k}]^2_{t,t} \\
&&\hspace{2cm}+\beta^{-2}_{jk}
\sigma^4 \sum\limits_{t<t'}\{[\bS_{k}]^2_{t,t'}+[\bS_{k}]^2_{t',t}+2[\bS_{k}]_{t,t'}[\bS_{k}]_{t',t}\}+\beta^{-2}_{jk}
\sigma^4 \Big(\sum\limits_{t=2}^T  [\bS_{k}]_{t,t}\Big)^2.
\endn
Thus its variance is such that
\beginy
\Big(4\beta^{-2}_{jk}+\sum\limits_{j'\ne j} \beta^{-2}_{j'k}\Big)\sigma^4R(K)h^{-1}T^{-2}\sum\limits_{l=1}^{T-1}(T-l)a_{l;k}.\label{hu}
\endy
From (\ref{tong}) and (\ref{hu}), we could deduce that
 $\beps^\top_{j}\delta\bbeta_j$ has   mean  of $pK(0)\sigma^2h^{-1}$ and variance
\beginn
&&\sigma^4R(K)h^{-1}T^{-2}\sum\limits_{k=1}^p\{4+\sum\limits_{j'\ne j} (\beta_{jk}/\beta_{j'k})^2\}
\sum\limits_{l=1}^{T-1}(T-l)a_{k,l}\\
&&\hspace{2cm} +\sigma^4T^{-2}\sum\limits_{k\ne k'}\{4+\sum\limits_{j'\ne j} (\beta_{jk}/\beta_{j'k})^2\}\sum\limits_{l=1}^{T-1}(T-l)b_{l;k,k'}.
\endn
Under assumption [A4], the variance of $\beps^\top_{j}\delta\bbeta_j$ could be further simplified as 
\beginn
&&\sigma^4R(K)h_k^{-1}\sum\limits_{k=1}^p c_k\{4+\sum\limits_{j'\ne j} (\beta_{jk}/\beta_{j'k})^2\}.
\endn
Now we deal with the fourth term in (\ref{divide}).
As the $k$th element of  $\beps^\top_{j}\delta$ given by
\beginn
\sum\limits_{j'=1}^n\beta^{-1}_{j'k}\Big(\sum\limits_{t=2}^T \epsilon_{tj}\gamma_{tk}^\top\Big) \bS_{k}\beps_{j'}
=\sum\limits_{j'\ne j} \beta^{-1}_{j'k}\Big(\sum\limits_{t,t'=2}^T \epsilon_{tj}\epsilon_{t'j'} [\bS_{k}]_{t,t'} \Big)+
 \beta^{-1}_{jk}\Big(\sum\limits_{t,t'=2}^T \epsilon_{tj}\epsilon_{t'j} [\bS_{k}]_{t,t'} \Big),
\endn
and  the $k'$th element of  $\bg^\top \beps_{j}$ given by $\sum\limits_{t=1}^T \frac{x_{tk'}}{x_{1k'}}\epsilon_{tj}$, we have
\beginn
&&\beps_{j}^\top \delta({\bg}^\top {\bg})^{-1} \bg^\top \beps_{j}
=\frac{1}{T}\sum\limits_{k,k'=1}^p \sigma_{k,k'}\beta^{-1}_{jk}
\Big(\sum\limits_{t=1}^T \frac{x_{tk'}}{x_{1k'}}\epsilon_{tj}\Big)\Big(\sum\limits_{t,t'=2}^T \epsilon_{tj}\epsilon_{t'j} [\bS_{k}]_{t,t'} \Big)\\
&&\hspace{3.3cm}+\frac{1}{T}\sum\limits_{k,k'=1}^p \sigma_{k,k'}\Big(\sum\limits_{t=1}^T \frac{x_{tk'}}{x_{1k'}}\epsilon_{tj}\Big)
\sum\limits_{j'\ne j} \beta^{-1}_{j'k}\Big(\sum\limits_{t,t'=2}^T \epsilon_{tj}\epsilon_{t'j'} [\bS_{k}]_{t,t'} \Big),
\endn
which is of mean zero with its variance  easily shown to be of order $O((Th)^{-1})$.

That $\beps^\top_{j}\delta$ is the dominating term in the partition (\ref{mao}) of    $\RSS_{0,j}-\RSS_{1,j}$, applies to any $j=1,\dots,p$. 
To  derive the asymptotics of $\lambda(H_0)$, we also need to consider the   
 covariance between $\RSS_{0,j}-\RSS_{1,j}$ and $\RSS_{0,j'}-\RSS_{1,j'}$
  ($j,{\tilde j}=1,\cdots,n, \ j\ne \tilde j$). This in turn equals to that between
   $\beps^\top_{j}\delta\bbeta_j$ and $\beps^\top_{\tilde j}\delta\bbeta_{\tilde j}$ 
, which is easily seen to be given by
\beginn
h^{-1}\sigma^4R(K) T^{-2}\sum\limits_{k=1}^p \sum\limits_{l=1}^{T-1}(T-l)a_{l;k}.
\endn
The proof is thus complete. $\hspace{\fill}\square$

 {\bf Proof of Corollary \ref{c1}}
 For backfitting estimation of additive models,  Opsomer (2000) studied theoretical properties on general linear smoothers with independent observations.
 We now describe the extension of his results to our case, i.e. for any given $j = 1, \ \cdots,  n$,   the estimation of $\{G_{jk}(.), k=1,\cdots, p\}$ 
 based on time series data $\{r_{tj},\, t=1,\cdots,T\}$ .

With linear smoother  matrices such as the $T\times T$ matrices $\bS_{k}$ , $k=1,\cdots,p$ of (\ref{smooth}), the backfitting estimates of the additive component functions evaluated at the observation points are by definition the solution to the following system of equations for the unknown vectors of fits $\bG_{j1},\cdots,\bG_{jp}$:
\beginy
\left[
\begin{array}{cccc}
\bI &\bS_{1}&\cdots&\bS_{1}\\
\bS_{2}&\bI&\cdots&\bS_{2}\\
\vdots&\vdots&\ddots&\vdots\\
\bS_{p}&\bS_{p}&\cdots&\bI
\end{array}\right]
\left[
\begin{array}{c}
\bG_{j1}\\\bG_{j2}\\ \vdots\\ \bG_{jp}
\end{array}\right]
=\left[
\begin{array}{c}
\bS_{1}\\\bS_{2}\\ \vdots\\ \bS_{p}
\end{array}\right] \bR_j.\label{fenghe}
\endy
Conceptually the solution  could be written  as
\beginy
\left[
\begin{array}{c}
\hat{\bG}_{j1}\\
\hat{\bG}_{j2}\\ \vdots\\ \hat{\bG}_{jp}
\end{array}\right]
=\left[
\begin{array}{cccc}
\bI &\bS_{1}&\cdots&\bS_{1}\\
\bS_{2}&\bI&\cdots&\bS_{2}\\
\vdots&\vdots\ddots&\vdots\\
\bS_{p}&\bS_{p}&\cdots&\bI
\end{array}\right]^{-1}
\left[
\begin{array}{c}
\bS_{1}\\\bS_{2}\\ \vdots\\ \bS_{p}
\end{array}\right] 
\bR_j\equiv \bM^{-1}
\bC \bR_j,\label{hua}
\endy
provided that $\bM$ is invertible. Write
\beginn
\bW_{k}=\bE_k \bM^{-1}\bC,
\endn
where $\bE_k$ is a partitioned matrix of dimension $T\times (T\, p)$ with an $T\times T$ identity matrix as the $k$th block and zero matrices else where, so that $\hat{\bG}_{jk}=\bW_{k}R_j.$ According to Lemma 2.1 of Opsomer (2000), equation (\ref{fenghe}) solved through backfitting algorithm will converge to a unique solution if
\beginy
\|\bS_{k}\bW_{[-k]}\|<1\label{guardian}
\endy
for some $k\in \{1,\cdots,p\}$ and any matrix norm $\|.\|$, where recall that $\bW_{[-k]}$ has been defined preceding Corollary \ref{c1}. 
As pointed out in Buja et al. (1989) and Opsomer (2000), a necessary condition for (\ref{guardian}) to hold for any of the major smoothing techniques unless the smoother matrices are centered, i.e.  $\bS_{k}$ replaced by its centered counterpart  $\bS_k^* $.
In that case, the additive smoother with respect to the $k$th component function $G_{jk}(.)$ is written as
\beginy
\bW_{k}=\bI-(\bI-\bS^*_{k}\bW_{[-k]})^{-1}(\bI-\bS^*_{k})=(\bI-\bS^*_{k}\bW_{[-k]})^{-1}\bS^*_{k}(\bI-\bW_{[-k]}). \label{hitchens}
\endy
 The aymptotic bias and variance of $\hat{\bG}_{jk},\, j=1,\cdots,T,\, k=1,\cdots, P$ is then derived from (\ref{hitchens}) and that $\hat{\bG}_{jk}=\bW_{k}R_j $; see Theorem 3.1 in Opsomer (2000) in the case of iid observations. Here we need to generalize these results to dependent sequences. The key intermediary step is, as in Opsomer  and Ruppert  (1997) and Opsomer (2000, pp. 178), to  show that
   that
 \beginn
&& \bS_k^*=\bS_k-{\bf 1}_{\T}{\bf 1}_{\T}^\top /T+o_p({\bf 1}_{\T}{\bf 1}_{\T}^\top /T),\\
&& (\bI-\bS_k^*\bW_{[-k]})^{-1}=\bI+O_p({\bf 1}_{\T}{\bf 1}_{\T}^\top /T),
 \endn
 uniformly over all elements of the matrices. This follows from   results given in Yu (1994) on rates of convergence for empirical processes of stationary mixing. The rest of the proof are identical to that of Theorem 3.1 of Opsomer (2000).
 $\hspace{\fill}\square$


\end{document}